\documentclass[12pt]{article}
\pdfoutput=1 
\usepackage{tikz}
\usepackage{amssymb}
\usepackage{epsfig}
\usepackage{bbm}
\usepackage{bbold}
\usepackage{graphicx}
\usepackage{hyperref}
\usepackage{romannum}
\usepackage{graphics,graphpap}
\usepackage{amsmath,exscale,amsthm}
\usepackage{yfonts}
\usepackage{mathrsfs}
\usepackage{float}
\usepackage{hyperref}
\usepackage{subcaption}

\usepackage[none]{hyphenat}
\usetikzlibrary{matrix,positioning,decorations.pathreplacing}

\pdfoutput=1
\usepackage{color}
\usepackage{amssymb}
\usepackage{epsfig}
\usepackage{romannum}

\def\e{\varepsilon}

\newcommand{\wt}{\widetilde}

\usepackage{amsmath}
\usepackage{amsfonts}

\begin{document}

\def\a{\alpha}
\def\b{\beta}
\def\c{\chi}
\def\d{\delta}
\def\e{\epsilon}
\def\f{\phi}
\def\g{\gamma}
\def\h{\eta}
\def\i{\iota}
\def\j{\psi}
\def\k{\kappa}
\def\la{\lambda}
\def\m{\mu}
\def\n{\nu}
\def\o{\omega}
\def\p{\pi}
\def\q{\theta}
\def\r{\rho}
\def\s{\sigma}
\def\t{\tau}
\def\u{\upsilon}
\def\x{\xi}
\def\z{\zeta}
\def\D{\Delta}
\def\F{\Phi}
\def\G{\Gamma}
\def\J{\Psi}
\def\L{\Lambda}
\def\O{\Omega}
\def\P{\Pi}
\def\Q{\Theta}
\def\S{\Sigma}
\def\U{\Upsilon}
\def\X{\Xi}

\def\ve{\varepsilon}
\def\vf{\varphi}
\def\vr{\varrho}
\def\vs{\varsigma}
\def\vq{\vartheta}

\def\dg{\dagger}                                     
\def\ddg{\ddagger}                                   
\def\wt#1{\widetilde{#1}}                    
\def\mt{\widetilde{m}_1}
\def\mti{\widetilde{m}_i}
\def\rt{\widetilde{r}_1}
\def\mtt{\widetilde{m}_2}
\def\mttt{\widetilde{m}_3}
\def\rtt{\widetilde{r}_2}
\def\mb{\overline{m}}
\def\VEV#1{\left\langle #1\right\rangle}        
\def\be{\begin{equation}}
\def\ee{\end{equation}}
\def\ds{\displaystyle}
\def\ra{\rightarrow}

\def\bea{\begin{eqnarray}}
\def\eea{\end{eqnarray}}
\def\NO{\nonumber}
\def\Bar#1{\overline{#1}}


\def\pl#1#2#3{Phys.~Lett.~{\bf B {#1}} ({#2}) #3}
\def\np#1#2#3{Nucl.~Phys.~{\bf B {#1}} ({#2}) #3}
\def\prl#1#2#3{Phys.~Rev.~Lett.~{\bf #1} ({#2}) #3}
\def\pr#1#2#3{Phys.~Rev.~{\bf D {#1}} ({#2}) #3}
\def\zp#1#2#3{Z.~Phys.~{\bf C {#1}} ({#2}) #3}
\def\cqg#1#2#3{Class.~and Quantum Grav.~{\bf {#1}} ({#2}) #3}
\def\cmp#1#2#3{Commun.~Math.~Phys.~{\bf {#1}} ({#2}) #3}
\def\jmp#1#2#3{J.~Math.~Phys.~{\bf {#1}} ({#2}) #3}
\def\ap#1#2#3{Ann.~of Phys.~{\bf {#1}} ({#2}) #3}
\def\prep#1#2#3{Phys.~Rep.~{\bf {#1}C} ({#2}) #3}
\def\ptp#1#2#3{Progr.~Theor.~Phys.~{\bf {#1}} ({#2}) #3}
\def\ijmp#1#2#3{Int.~J.~Mod.~Phys.~{\bf A {#1}} ({#2}) #3}
\def\mpl#1#2#3{Mod.~Phys.~Lett.~{\bf A {#1}} ({#2}) #3}
\def\nc#1#2#3{Nuovo Cim.~{\bf {#1}} ({#2}) #3}
\def\ibid#1#2#3{{\it ibid.}~{\bf {#1}} ({#2}) #3}

\title{
\vspace*{1mm}
{\bf The $SO(10)$-inspired leptogenesis \\ timely opportunity}
\author{
{\Large Pasquale Di Bari and  Rome Samanta}
\\
{\it Physics and Astronomy}, 
{\it University of Southampton,} \\
{\it  Southampton, SO17 1BJ, U.K.} 
\\
}}
\maketitle \thispagestyle{empty}
\pagenumbering{arabic}

\begin{abstract}
We study the connection between absolute neutrino mass and neutrino mixing parameters within $SO(10)$-inspired leptogenesis.
We show that current favoured values of the unknown neutrino mixing parameters  point toward values of
the absolute neutrino mass scale that will be fully tested by cosmological observations and neutrinoless 
double beta decay experiments during next years.  In particular, for $m_{D2}/m_{\rm charm} \leq 5$, 
where $m_{D2}$ is the intermediate Dirac neutrino mass, and for current best fit values of the Dirac phase $\delta$ and 
the atmospheric mixing angle $\theta_{23}$, 
we derive a lower bound  on the neutrinoless double beta decay effective neutrino mass 
$m_{ee} \gtrsim 31\,{\rm meV}$ and on the sum of the neutrino masses $\sum_i m_i \gtrsim 125\,{\rm meV}$.
These lower bounds hold for normally ordered neutrino masses,  as currently favoured by global analyses, and approximately
for $\delta \in [155^\circ,240^\circ]$  and $\theta_{23}$ in the second octant.  
If values in this region will be confirmed by future planned long baseline experiments,
then  a signal at next generation  neutrinoless double beta decay experiments 
is expected, despite neutrino masses being  normally ordered. 
Outside this region, the lower bounds strongly relax but a great fraction of the allowed range of values 
still allows a measurement of the lightest neutrino mass. Therefore, in the the next years low energy neutrino experiments will provide a stringent test of $SO(10)$-inspired leptogenesis, that might result  either in severe constraints or in a strong evidence. 
\end{abstract}

\newpage
\section{Introduction}

The persistent lack of evidence of new physics at colliders 
supports the idea that the  matter-antimatter asymmetry of the Universe originates 
from a dynamical process occurred during the early history of the universe
at energies well above the electroweak energy scale.  
From this point of  view minimal scenarios of leptogenesis,  relying  on  type-I seesaw mechanism \cite{seesaw} for the generation
of neutrino masses and mixing and on the assumption of thermal leptogenesis \cite{fy}, are very attractive. 
 The asymmetry is generated at an energy scale approximately corresponding 
 to the mass of the right-handed (RH) neutrino species whose decays generate the asymmetry.
 Barring fine tuned  solutions and unnaturally low neutrino Yukawa couplings, 
this mass has to be very high in order  to reproduce the solar and 
 atmospheric neutrino mass scales measured in neutrino oscillation experiments.
This is  in nice agreement with the lower bound $M_I \gtrsim 10^{9}\, {\rm GeV}$
on the mass of the heavy neutrino producing the asymmetry obtained imposing 
successful leptogenesis \cite{di,cmb,flavourlep}. 
However,  the possibility to test such very high energy scale leptogenesis scenarios necessarily relies 
on some strategy to reduce the number of independent parameters in the type-I seesaw mechanism. 
In this respect an attractive  way to realise such a reduction is provided by
$SO(10)$-inspired conditions \cite{SO10inspired} since they are naturally 
satisfied in various (not necessarily $SO(10)$) grand-unified models.

The resulting RH neutrino mass spectrum is very hierarchical and typically for
the lowest mass one has $M_1 \sim 10^5\,{\rm GeV}$, certainly well below the lower 
bound $10^9\,{\rm GeV}$, in a way that the asymmetry produced by its decays is negligible.
However, the next-to-lightest RH neutrino typically has a mass
$M_{2}\sim (10^{10}$--$10^{11})\,{\rm GeV}$, nicely in the right range 
for its flavoured $C\!P$  asymmetries to be sufficiently large to attain successful leptogenesis.
In this way, one is necessarily led  to consider $N_{2}$-leptogenesis, where 
the observed asymmetry is reproduced by the next-to-lightest RH neutrino decays \cite{geometry}.
Within the $SO(10)$-inspired leptogenesis scenario \cite{riotto1}
the asymmetry can be expressed, in first approximation, as a function of the nine 
low energy neutrino parameters and just one Dirac neutrino mass. The latter is 
constrained by $SO(10)$-inspired conditions to be not too different by the
charm quark mass. In this way the successful leptogenesis condition generates constraints 
in the space of all nine low energy neutrino parameters \cite{riotto1,riotto2}. 
  Low energy neutrino phases are particularly constrained since they 
  play an important direct role both in maximising the asymmetry produced
by $N_2$-decays and in making possible for this to escape the lightest RH neutrino 
  wash-out from inverse processes \cite{decrypting}.

The most interesting constraint is a lower bound on the 
lightest neutrino mass, $m_1 \gtrsim {\cal O}(1)\,{\rm meV}$ \cite{riotto1}.
This lower bound also translates into a lower bound on the neutrinoless double beta deday
effective neutrino mass. In general, it is well known that for normally ordered neutrino mass this can be 
arbitrarily small if $m_1$ is approximately within the range $(3$--$7)\,{\rm meV}$. However, within $SO(10)$-inspired
leptogenesis one has  a lower bound $m_{ee} \gtrsim {\cal O}(0.1)\,{\rm meV}$ \cite{full}.
The existence of such a lower bound on the absolute neutrino mass scale is very interesting, 
since it represents  a strong constraint on any $SO(10)$-inspired model that aims at 
embedding successful leptogenesis. However,
no current or planned absolute neutrino mass scale experiment has the sensitivity to fully test 
such a lower bound in a way either to rule out $SO(10)$-inspired leptogenesis or to 
measure a value of the absolute neutrino mass scale in agreement with the lower bound. 
On the other hand, a positive signal in neutrinoless 
double beta decay experiments, at the level of $m_{ee} \sim 10\,{\rm meV}$, 
would certainly represent a strong support to $SO(10)$-inspired 
leptogenesis, since it would first of all establish lepton number violation,  a fundamental ingredient for leptogenesis
models,  and it would fit very well with the expectations for the bulk of solutions.

An analytical expression of the lower bound was first derived  neglecting
the mismatch between the neutrino and charged lepton flavour basis \cite{decrypting}. 
When this is taken into account,  scatter plots show that the lower bound on the lightest 
neutrino mass gets slightly relaxed  \cite{riotto2,full}.  The dependence of the lower bound on the
Dirac phase delta and on the atmospheric mixing angle was separately (i.e., marginalising on one of the two) 
studied in \cite{full}  and, interestingly, the results clearly showed that the lower bound is modulated by the 
value of the Dirac phase and can become much more stringent away from $\d = 2\,n\,\pi$ (with $n$ integer).  
Moreover, it becomes more and more stringent also for increasing values of the atmospheric mixing angle. 
Interestingly, latest results from neutrino oscillation experiments go in this direction for both parameters,  
thus favouring a more stringent lower bound on the absolute neutrino mass scale.

In this paper we  study in detail how the lower bounds, on $m_1$ and on $m_{ee}$, 
jointly depend on both $\delta$ and $\theta_{23}$.
We show that for current best fit values, and approximately within $1\s$, 
the lower bound on $m_1$ is actually much more stringent, finding $m_1 \gtrsim 34\,{\rm meV}$
that corresponds to $\sum_i m_i \gtrsim 0.125\,{\rm eV}$. 
This lower bound is  already in slight tension with the upper bound from cosmological observations 
$\sum_i m_i < 0.146\,{\rm eV}$ ($95\%$ C.L.) \cite{hannestad}. 
Within the same region we also find $m_{ee} \gtrsim 31\,{\rm meV}$, a lower bound that, quite interestingly, 
will be tested by next generation $0\nu\b\b$ experiments. 
    
These results  hold for $\a_2 \leq 5$, where $\a_2$ is the ratio of the
intermediate Dirac neutrino mass to the charm quark mass at the temperature of 
leptogenesis $T_{\rm lep} \simeq 5 \times 10^{10}\,{\rm GeV}$. 
As we will see, these lower bounds do not apply just for best fit values of $\delta$ and $\theta_{23}$ but 
for quite a large region in the plane $(\theta_{23},\delta)$, approximately for 
$\delta$ in the interval $155^\circ$--$240^\circ$   and for $\theta_{23}$ in the second octant. 
Outside this region the lower bound on $m_1$ drops quite sharply but
within a $2\s$ region around best fit values it is still much
more stringent than the lower bound
$m_1 \gtrsim 0.5 \, {\rm meV}$ that was found for $\delta = 2\,n\,\pi $ and $\theta_{23} \geq 36.5^\circ$  \cite{full}. 
Indeed as we will see the lower bound gets more stringent for increasing values of $\theta_{23}$ and, 
therefore, the fact that current data favour  $\theta_{23}$ in the second octant goes in that direction.

These results clearly show the strong  connection between absolute neutrino mass scale
and neutrino mixing parameters within $SO(10)$-inspired leptogenesis,  a connection that is quite a
distinguished feature of the scenario and that is the main focus of our investigation.

The paper is organised as follows. In Section 2 we review how within type-I seesaw mechanism one 
can impose $SO(10)$-inspired conditions and  reproduce the matter-antimatter asymmetry of the universe
with $SO(10)$-inspired leptogenesis.  We also briefly review current experimental results on neutrino masses and mixing parameters.
In Section 3 we show the results of scatter plots projected on the 3-dim spaces $(\delta, \theta_{23}, m_1)$ and 
$(\delta, \theta_{23}, m_{ee})$. The 3-dim projections of the scatter plots show clearly how $SO(10)$-inspired leptogenesis
identifies a special region strongly connecting the absolute neutrino mass to $\delta$ and $\theta_{23}$.
In Section 4 we focus on the lower bound on the absolute neutrino mass scale that can be extracted from these
scatter plots. We show the lower bounds on $m_1$ and  $m_{ee}$, in the form of isocontour lines in the plane $(\theta_{23}, \delta)$
for $\alpha_2  = 5$ and for a misalignment between the neutrino Yukawa basis and the charged lepton flavour basis no larger  than the one measured in the quark sector and encoded by the CKM matrix. 
In Section 5 we show the dependence of the lower bounds on $\alpha_2$  and, more generally, 
on the exact definition of $SO(10)$-inspired conditions. 
In particular, we show how the lower bounds get progressively relaxed allowing for larger 
and larger values of the angles parameterising 
the left-handed leptonic mixing matrix describing the mismatch between 
neutrino Yukawa and charged lepton flavour basis,  the analogue of the CKM matrix.  
In Section 6 we derive an analytical expression for the lower bound on $m_1$
applying the analytical procedure discussed in \cite{decrypting} for $V_L = I$ and in \cite{full} for $0 \leq V_L \leq V_{CKM}$. 
This expression clearly shows the dependence on $\theta_{23}$ and $\delta$.  We also show analytically the effect played by
taking $V_L \simeq V_{CKM}$ of allowing a complete suppression of the lightest RH neutrino wash-out.
Finally, in Section 7 we draw conclusions, discussing in particular how absolute neutrino mass experiments 
might have the opportunity in next years, depending on the results on $\delta$ and $\theta_{23}$, either to rule out or to find quite 
a strong signature of $SO(10)$-inspired leptogenesis, considering the interplay between 
absolute neutrino mass and neutrino mixing parameters.


 \section{Neutrino masses and $SO(10)$-inspired leptogenesis}   %

The $SO(10)$-inspired leptogenesis scenario relies, in its minimal form, 
on the assumption that neutrino masses and mixing are described
by the type-I seesaw mechanism with three RH neutrinos. 
The light neutrino mass matrix is then given by the seesaw formula \cite{seesaw}
\be
m_{\nu} = - m_D \, {1\over D_M} \, m_D^T  \,   .
\ee
Here we indicated with $m_D$ the neutrino Dirac mass matrix in the flavour basis, where both charged lepton and Majorana mass matrices are diagonal, and defined $D_M \equiv {\rm diag}(M_1, M_{2}, M_{3})$, where $M_1 \leq M_{2} \leq M_{3}$, 
are the three heavy neutrino masses.  
In the flavour basis, the light neutrino mass matrix  is diagonalised by the leptonic mixing matrix $U$, 
in a way that the light neutrino masses $m_1 \leq m_2 \leq m_3$ are given by 
\be
D_m = - U^{\dagger}\,m_{\nu}\,U^\star \,   ,
\ee
where $D_m \equiv {\rm diag}(m_1,m_2,m_3)$.  Neutrino oscillation experiments measure
the atmospheric neutrino mass scale $m_{\rm atm} \equiv \sqrt{m^2_3 - m^2_1} = (49.9 \pm 0.3)\,{\rm meV}$
and the solar neutrino mass scale $m_{\rm sol} \equiv \sqrt{m^2_2 - m^2_1} = (8.6 \pm 0.1) \,{\rm meV}$ \cite{nufit}. We consider only normally ordered neutrino masses since the case of inverted ordering is not only  disfavoured  by current data at $\sim 3\s$,  
but also only marginally viable in $SO(10)$-inspired leptogenesis.
As  mentioned in the introduction, cosmological observations place a stringent upper bound 
$\sum_i m_i < 0.146\,{\rm eV}$ ($95\%$ C.L.) \cite{hannestad} on the sum of neutrino masses
for normally ordered neutrino masses, corresponding to an upper bound $m_1 < 43\,{\rm meV}$ ($95\%$ C.L.).

The leptonic mixing matrix can then be parameterised
in terms of the usual mixing angles $\theta_{ij}$, the Dirac phase $\d$
and the Majorana phases $\rho$ and $\s$, 
\be
U=  \left( \begin{array}{ccc}
c_{12}\,c_{13} & s_{12}\,c_{13} & s_{13}\,e^{-{\rm i}\,\d} \\
-s_{12}\,c_{23}-c_{12}\,s_{23}\,s_{13}\,e^{{\rm i}\,\d} &
c_{12}\,c_{23}-s_{12}\,s_{23}\,s_{13}\,e^{{\rm i}\,\d} & s_{23}\,c_{13} \\
s_{12}\,s_{23}-c_{12}\,c_{23}\,s_{13}\,e^{{\rm i}\,\d}
& -c_{12}\,s_{23}-s_{12}\,c_{23}\,s_{13}\,e^{{\rm i}\,\d}  &
c_{23}\,c_{13}
\end{array}\right)
\, {\rm diag}\left(e^{i\,\rho}, 1, e^{i\,\sigma}
\right)\,   .
\ee
Latest  global analyses of neutrino oscillation experiment results find, in the case of normal ordering,
the following best fit values, $1\s$ errors  and $3\s$ intervals
 for the mixing angles and the  leptonic Dirac phase $\d$  \cite{nufit}: 
\bea\label{expranges}
\theta_{13} & = &  8.60^{\circ}\pm 0.13^{\circ} \in [8.22^{\circ}, 8.98^{\circ}] \,  ,  \label{theta13} \\ 
\theta_{12} & = &  33.82^{\circ}\pm 0.76^{\circ} \in [31.61^{\circ}, 36.27^{\circ}]  \,  ,  \\ 
\theta_{23} & = &  {48.6^{\circ}}^{+1.0^{\circ}}_{-1.4} \in [40.8^{\circ}, 51.3^{\circ}]  \,  ,  \label{theta23}  \\ 
\d & = &  {222^{\circ}} ^{+39^{\circ}}_{-28^{\circ}} \in  [144^{\circ}, 357^{\circ}]  \, .  \label{deltarange}
\eea 
At the moment, not only there are no experimental constraints on the Majorana phases, but we do not even know
whether they are physical, in the case of Majorana neutrinos, or unphysical, in the case of Dirac neutrinos.

The type-I seesaw extension of the SM, introduces eighteen additional parameters
and predicts that neutrinos are Majorana particles. On the other hand
low energy neutrino experiments can measure only eight independent parameters, 
including the experimental information on  the effective neutrinoless double beta decay neutrino mass
\be\label{mee}
m_{ee} \equiv |m_{\nu ee}| = \left|m_1\,U^2_{e1}+m_2\,U^2_{e2}+m_3\,U^2_{e3} \right| \,   ,
\ee
coming from $0\nu\beta\beta$ experiments. Having not found a positive signal so far, 
they place an upper bound with the most stringent one coming from
the KamLAND-ZEN experiment that finds $m_{ee}  < 165\,{\rm meV} \, (90\%\,{\rm C.L.})$ \cite{kamlandzen}.    
Therefore, the type-I seesaw mechanism cannot be tested in a model independent way and, to this extent, 
one needs to introduce some additional (phenomenological and/or theoretical) information to reduce the number 
of independent parameters.  The $SO(10)$-inspired leptogenesis scenario is a  well justified framework 
that realises such a reduction, yielding testable experimental predictions on low energy neutrino
parameters. 

Qualitatively,  $SO(10)$-inspired conditions are equivalent to the assumption that 
the neutrino Dirac mass matrix  is not {\em too different} from the up quark mass matrix $m_u$. 
This is a property that is certainly realised in $SO(10)$ models \cite{SO10models} but 
in recent years also non-$SO(10)$ models respecting $SO(10)$-inspired conditions 
have been proposed \cite{king}.  If one considers $SO(10)$ models, fermion families are represented
by 16-dim spinors of $SO(10)$. In the simplest case, the dominant contribution to the  Yukawa coupling matrices  
comes from the 10-dim Higgs multiplet. In this case one would simply have $m_D = m_u = m_d = m_{\ell}$, where 
$m_u$, $m_d$ and $m_{\ell}$ are respectively the up quark, down quark and charge lepton mass matrices, and
there would be no mixing whatsoever, neither in the quark sector nor in the lepton sector. 
For this reason, in order to get realistic models, one has to add some higher dimensional Higgs multiplet 
that introduces a mismatch among fermion matrices and is responsible for the observed leptonic and quark mixing. 
In the case of $SO(10)$ models, contributions from 120-dim and 126-dim Higgs multiplets 
introduce in general such kind of mismatch and can indeed successfully reproduce the mixing
both in the lepton and quark sectors \cite{SO10models,babu}.  In our case, to be more general and following \cite{riotto1,riotto2}, 
we define $SO(10)$-inspired models that class of models that satisfy  $SO(10)$-inspired conditions defined as follows.

Let us parameterise the neutrino Dirac mass  matrix in the bi-unitary parameterisation,
\be\label{biunitary}
m_D = V_L^\dagger \, D_{m_D} \, U_R \,   ,
\ee
where $V_L$ is the unitary matrix acting on left-handed neutrino fields and realising the transformation from the flavour basis 
to the neutrino Yukawa basis (where $m_D$ is diagonal instead of the charged lepton mass matrix). 
It is then the analogous of the CKM matrix in the quark sector,
encoding the mismatch between the neutrino Yukawa basis  and the charged lepton flavour basis. It can be 
parameterised analogously to the leptonic mixing matrix as
\begin{equation}\label{VLmatrix}
V_L=
\left( \begin{array}{ccc}
c^L_{12}\,c^L_{13} & s^L_{12}\,c^L_{13} & s^L_{13}\,e^{-{\rm i}\,\d_L} \\
-s^L_{12}\,c^L_{23}-c^L_{12}\,s^L_{23}\,s^L_{13}\,e^{{\rm i}\,\d_L} &
c^L_{12}\,c^L_{23}-s^L_{12}\,s^L_{23}\,s^L_{13}\,e^{{\rm i}\,\d_L} & s^L_{23}\,c^L_{13} \\
s^L_{12}\,s^L_{23}-c^L_{12}\,c^L_{23}\,s^L_{13}\,e^{{\rm i}\,\d_L}
& -c^L_{12}\,s^L_{23}-s^L_{12}\,c^L_{23}\,s^L_{13}\,e^{{\rm i}\,\d_L}  &
c^L_{23}\,c^L_{13}
\end{array}\right)
\, {\rm diag}\left(e^{i\,\rho_L}, 1, e^{i\,\sigma_L}  \right)\, ,
\end{equation}
in terms of three mixing angles $\theta_{12}^L, \theta_{13}^L$ 
and $\theta_{23}^L$ ($s^L_{ij} \equiv \sin \theta_{ij}^L$ and $c_{ij} \equiv \cos\theta_{ij}^L$),
one Dirac-like phase $\delta_L$ and two Majorana-like phases $\rho_L$ and $\sigma_L$.
The diagonal matrix $D_{m_D} \equiv {\rm diag}(m_{D1}, m_{D2}, m_{D3})$ gives
the spectrum of Dirac neutrino masses. Finally, $U_R$ acts on the RH neutrino fields and it is the matrix encoding the mismatch between
the neutrino Yukawa basis and the flavour basis, where  the Majorana mass matrix is diagonal. It can then be regarded
as the RH neutrino mixing matrix. 
We define $SO(10)$-inspired models that class of models respecting the following $SO(10)$-inspired conditions: 
\begin{itemize}
\item[i)] The unitary matrix $V_L$ has mixing angles $0\leq \theta_{ij}^L \leq \theta_{ij}^{CKM}$, 
where $\theta_{ij}^{CKM}$ are the mixing angles in the CKM matrix and, in particular, 
$\theta_{12}^{CKM}\simeq 13^{\circ}$ is the Cabibbo angle; 
\item[ii)] The neutrino Dirac masses are such that the 
mass ratios $\a_1 \equiv m_{D1}/m_{\rm up}$, $\a_2 \equiv m_{D2}/m_{\rm charm}$, 
$\a_3 \equiv m_{D3}/m_{\rm top}$ are ${\cal O}(1)$ parameters, 
more precisely we allow them to vary within $[0.1, 10]$. 
Notice that the up quark masses have to be evaluated at the energy scale of interest.
In our case we are interested in temperatures where the asymmetry is generated, i.e., for
$T \sim (10^{10}$--$10^{11})\,{\rm GeV}$ and we will take 
$m_{\rm up}=1\,{\rm MeV}$, $m_{\rm charm}=400\,{\rm MeV}$ and $m_{\rm top}= 100\,{\rm GeV}$ 
\cite{fusaokakoide}.
\end{itemize}
With these assumptions the three RH neutrino masses are well expressed
in terms of $m_{\nu}$, the three $\a_i$'s and $V_L$ by \cite{full}
\be\label{MI}
M_1    \simeq   {m^2_{D1} \over |\widetilde{m}_{\nu 11}|} \, , \;\;
M_{2}  \simeq    {m^2_{D2} \over m_1 \, m_2 \, m_3 } \, {|\widetilde{m}_{\nu 11}| \over |(\widetilde{m}_{\nu}^{-1})_{33}|  } \,  ,  \;\;
M_{3}  \simeq   m^2_{D3}\,|(\widetilde{m}_{\nu}^{-1})_{33}|  ,
\ee
where $\widetilde{m}_{\nu} \equiv V_L\,m_{\nu}\,V_L^T$ is the light neutrino mass matrix in the 
neutrino Yukawa basis.  The resulting spectrum of neutrino masses is very hierarchical and in particular
one has $M_1 \ll 10^{9}\,{\rm GeV}$ and $M_{2}/M_{3} \ll 1$\footnote{We are barring the very fine tuned compact spectrum 
solution with $M_1 \sim M_2 \sim M_3$ \cite{compact}.}. This results into a negligible contribution to the
final matter-antimatter asymmetry produced by $N_1$- and $N_{3}$-decays, so that the only contribution
that can reproduce the observed asymmetry comes from $N_{2}$-decays. The final $B-L$ asymmetry, that is
conserved in the standard model and in particular by sphaleron processes,  has to be calculated
as the sum of three charged lepton flavour asymmetries ($\a = e,\mu, \tau$)
\be\label{NBmLf}
N_{B-L}^{\rm f} = \sum_\a \, N_{\D_\a} \,  ,
\ee
where $\D_{\a} \equiv B/3 - L_{\a}$. A fraction $a_{\rm sph}=28/79$ of the final $B-L$ asymmetry
will ultimately be in the form of a baryon asymmetry at the sphaleron freeze-out time. In this way the 
baryon-to-photon ratio predicted by leptogenesis can be calculated as 
\be\label{etaBlep}
\eta_B^{\rm lep} =a_{\rm sph}\,{N_{B-L}^{\rm f}\over N_{\g}^{\rm rec}} \simeq 
0.96\times 10^{-2}\,N_{B-L}^{\rm f} \,  ,
\ee
where $N_{\g}^{\rm rec}$ is the abundance of photons at recombination.   
The numerical expression on the right-hand side holds when the abundances are normalised in a way that the ultra-relativistic 
thermal equilibrium abundance of a RH neutrino species  is just given by 
$N^{\rm eq}_{N_I}(T \gg M_I) =1$. In this way the abundance of photons at recombination is given by
$N_{\g}^{\rm rec} = 4\, g_R^{\rm SM} /(3\,g_{S}^{\rm rec}) \simeq 36.4$, where $g_R^{\rm SM}=106.75$
is the number of standard model ultra-relativistic degrees of freedom and $g_{S}^{\rm rec} = 43/11$ is the
entropy ultra-relativistic number  of degrees of freedom at recombination. 

Notice that  we are assuming the contribution from a pre-existing asymmetry to be negligible. The possibility that a large
pre-existing asymmetry is generated by some external mechanism prior to leptogenesis and 
it is then washed-out by RH neutrinos inverse processes while decays produce the observed (much smaller) asymmetry, 
so-called {\em strong thermal leptogenesis} scenario, 
has been considered in \cite{problem} and it has been shown that this is possible
only within tauon-dominated $N_{2}$-leptogenesis with some additional conditions. 
Interestingly, this scenario can be realised within $SO(10)$-inspired leptogenesis 
leading to very sharp predictions on low energy neutrino parameters \cite{solution}.
Currently, the main challenge is the existence of an upper bound on the atmospheric mixing angle in tension with current
data favouring second octant. This can be reconciled only  for quite large values of $\a_2 \gtrsim 5$ \cite{chianese}
that, however, seem to be indicated also by realistic fits in $SO(10)$ models \cite{rodejohann}. 
In this paper we do not consider  strong thermal $SO(10)$-inspired leptogenesis scenario but we just remind that,
for this more restrictive scenario to be realised, there is anyway, independently of $\a_2$, a very stringent and 
compelling lower bound  on the absolute neutrino mass scale, with both $m_1$ and 
$m_{ee} \gtrsim 10\,(2)\,{\rm meV}$ \cite{sophie} for an initial pre-existing asymmetry $N^{\rm p,i}_{B-L}=10^{-1}\,(10^{-3})$. 
This is getting already tested by current cosmological observations while for neutrinoless double beta decay
signal we need to wait for next generation experiments.

The three flavoured  asymmetries  in Eq.~(\ref{NBmLf}) have to be calculated within 
$N_2$-leptogenesis \cite{geometry} and, taking into account also so-called phantom terms, 
these can be calculated using the expressions
\cite{vives,bounds,fuller,density}
\bea\label{twofl} \nonumber
N_{\D_e}^{\rm lep,f} & \simeq &
\left[{K_{2e}\over K_{2\tau_2^{\bot}}}\,\ve_{2 \tau_2^{\bot}}\kappa(K_{2 \tau_2^{\bot}}) 
+ \left(\ve_{2e} - {K_{2e}\over K_{2\tau_2^{\bot}}}\, \ve_{2 \tau_2^{\bot}} \right)\,\kappa(K_{2 \tau_2^{\bot}}/2)\right]\,
\, e^{-{3\pi\over 8}\,K_{1 e}}  \,   , \\ \nonumber
N_{\D_\m}^{\rm lep,f} & \simeq & \left[{K_{2\mu}\over K_{2 \tau_2^{\bot}}}\,
\ve_{2 \tau_2^{\bot}}\,\kappa(K_{2 \tau_2^{\bot}}) +
\left(\ve_{2\mu} - {K_{2\mu}\over K_{2\tau_2^{\bot}}}\, \ve_{2 \tau_2^{\bot}} \right)\,
\kappa(K_{2 \tau_2^{\bot}}/2) \right]
\, e^{-{3\pi\over 8}\,K_{1 \mu}} \,  , \\
N_{\D_\t}^{\rm lep,f} & \simeq & \ve_{2 \tau}\,\kappa(K_{2 \tau})\,e^{-{3\pi\over 8}\,K_{1 \tau}} \,  ,
\eea
that apply for $10^{9}\,{\rm GeV} \lesssim M_2 \lesssim 10^{12}\,{\rm GeV}$. 
In this mass range the asymmetry production occurs in the two fully flavoured
 regime \cite{flavoureffects}, where Boltzmann equations are used to describe the 
 evolution of the two flavoured asymmetries, the electronic and the sum of muonic and tauonic.\footnote{For $M_2 
 \gtrsim 10^{12}\, {\rm GeV}$ we should use a modified version where the production
occurs in the unflavoured regime. However,  in this case there are only very marginal solutions 
since the wash-out at the production is much stronger. We should also mention that we are 
neglecting flavour coupling. 
In \cite{flcoupling} it was noticed that including flavour coupling effects some new special 
solutions in a region where $\theta_{23}$ is deeply in the second octant ($\theta_{23}\simeq 53^\circ$) 
and $\delta \simeq 20^\circ$. We are not considering here these solutions since this region is not only disfavoured
by current data but also because those solutions imply quite a large fine tuning in the seesaw formula.}
 Moreover, it should also be noticed that they are valid for $M_1 \gtrsim T_{\rm sph}^{\rm out} \sim 100\,{\rm GeV}$, 
 since otherwise there would not be any wash-out from
 the lightest RH neutrino (an alternative scenario considered  in \cite{susy}). 
 This condition is  naturally realised for $\a_1 \gtrsim 0.1$, as
 we are assuming.  In Eqs.~(\ref{twofl}) the $\ve_{2\a}$'s are the $N_2$ flavoured $C\!P$ asymmetries 
 defined as $\ve_{2\a} \equiv -(\G_{2\a}-\overline{\G}_{2\a})/(\G_2 + \overline{\G}_2)$,
 where $\G_2 \equiv \sum_{\a} \G_{2\a}$ and $\overline{\G}_{2}\equiv \sum_{\a} \, \overline{\G}_{2\a}$ and
 we indicated with $\Gamma_{I\a}=\Gamma (N_I \ra \phi^\dagger \, l_\alpha)$ 
and $\bar{\Gamma}_{I \a}=\Gamma (N_I \ra \phi \, \bar{l}_\alpha)$ 
the zero temperature limit of the flavoured decay rates into $\a$ leptons
and anti-leptons respectively.
 
 Accounting for the interference between tree level and one loop graphs one obtains for the  
 flavoured $C\!P$ asymmetries \cite{crv}
\be\label{eps2a}
\ve_{2\a} \simeq
\overline{\ve}(M_2) \, \left\{ {\cal I}_{23}^{\a}\,\x(M^2_3/M^2_2)+
\,{\cal J}_{23}^{\a} \, \frac{2}{3(1-M^2_2/M^2_3)}\right\} \,  ,
\ee
where we introduced
\be
\overline{\ve}(M_2) \equiv {3\over 16\,\pi}\,{M_2\,m_{\rm atm} \over v^2} \, , 
\hspace{3mm} \xi(x)=\frac{2}{3}x\left[(1+x)\ln\left(\frac{1+x}{x}\right)-\frac{2-x}{1-x}\right] \,  ,
\ee
\be\label{Ial23}
{\cal I}_{23}^{\a} \equiv   {{\rm Im}\left[m_{D\a 2}^{\star}
m_{D\a 3}(m_D^{\dag}\, m_D)_{2 3}\right]\over M_2\,M_3\,\mtt\,m_{\rm atm} }\,   
\hspace{5mm}
\mbox{\rm and}
\hspace{5mm}
{\cal J}_{23}^{\a} \equiv  
{{\rm Im}\left[m_{D\a 2}^{\star}\, m_{D\a 3}(m_D^{\dag}\, m_D)_{3 2}\right] 
\over M_2\,M_3\,\mtt\,m_{\rm atm} } \,{M_2\over M_3}   ,
\ee
with $\mtt \equiv (m_D^{\dag}\, m_D)_{2 2}/M_2$.
Since $M_3 \gg M_2$,  one has $\xi(M_3^2/M_2^2) \simeq 1$
and the second term $\propto {\cal J}^{\a}_{23}$ can be neglected 
in Eq.~(\ref{eps2a}).  The expression (\ref{Ial23}) for the interference term ${\cal I}_{23}^{\a}$
can be recast using the bi-unitary parameterisation (Eq.~(\ref{biunitary})) and in this way
one obtains the following expression for the flavoured $C\!P$ asymmetries \cite{full}
\be\label{ve2alAN}
\ve_{2\a} \simeq {3 \over 16\, \pi \, v^2}\,
{|(\widetilde{m}_{\nu})_{11}| \over m_1 \, m_2 \, m_3}\,
{\sum_{k,l} \, m_{D k} \, m_{Dl}  \, {\rm Im}[V_{L k \a }  \,  V^{\star}_{L  l \a } \, 
U^{\star}_{R k 2}\, U_{R l 3} \,U^{\star}_{R 3 2}\,U_{R 3 3}] 
\over |(\widetilde{m}_{\nu}^{-1})_{33}|^{2} + |(\widetilde{m}_{\nu}^{-1})_{23}|^{2}}   \,  .
\ee
Like the three RH neutrino masses, the RH neutrino mixing matrix can be also expressed analytically in terms of 
$m_{\nu}$, the three $\a_i$'s and $V_L$. It is found \cite{full}
\be\label{UR}
U_R \simeq  
\left( \begin{array}{ccc}
1 & -{m_{D1}\over m_{D2}} \,  {\widetilde{m}^\star_{\nu 1 2 }\over \widetilde{m}^\star_{\nu 11}}  & 
{m_{D1}\over m_{D3}}\,
{ (\widetilde{m}_{\n}^{-1})^{\star}_{13}\over (\widetilde{m}_{\n}^{-1})^{\star}_{33} }   \\
{m_{D1}\over m_{D2}} \,  {\widetilde{m}_{\nu 12}\over \widetilde{m}_{\nu 11}} & 1 & 
{m_{D2}\over m_{D3}}\, 
{(\widetilde{m}_{\n}^{-1})_{23}^{\star} \over (\widetilde{m}_{\n}^{-1})_{33}^{\star}}  \\
 {m_{D1}\over m_{D3}}\,{\widetilde{m}_{\nu 13}\over \widetilde{m}_{\nu 11}}  & 
- {m_{D2}\over m_{D3}}\, 
 {(\widetilde{m}_\nu^{-1})_{23}\over (\widetilde{m}_\nu^{-1})_{33}} 
  & 1 
\end{array}\right) 
\,  D_{\Phi} \,  ,
\ee
where the three phases in 
$D_{\phi} \equiv {\rm diag}(e^{-i \, {\Phi_1 \over 2}}, e^{-i{\Phi_2 \over 2}}, e^{-i{\Phi_3 \over 2}})$ 
are given by
\be
\Phi_1 = {\rm Arg}[-\widetilde{m}_{\nu 11}^{\star}] \,  , \; \;
\Phi_2 = {\rm Arg}\left[{\widetilde{m}_{\nu 11}\over (\widetilde{m}_{\nu}^{-1})_{33}}\right] -2\,(\rho+\s)-2\,(\rho_L + \s_L) \, , \;  \;
\Phi_3 = {\rm Arg}[-(\widetilde{m}_{\nu}^{-1})_{33}] \,  .
\ee
With this analytical expression for the matrix $U_R$ and neglecting sub-dominant terms, one obtains the following
analytical expression for the tauonic $C\!P$ asymmetry depending only on $\a_2$, $m_{\nu}$ and $V_L$,
\be\label{ve2alANtau}
\ve_{2\tau}  \simeq  {3\,m^2_{D2} \over 16\, \pi \, v^2}\,
{|(\widetilde{m}_{\nu})_{11}| \over m_1 \, m_2 \, m_3}\,
{|(\widetilde{m}_{\nu}^{-1})_{23}| \over |(\widetilde{m}_{\nu}^{-1})_{33}|}\,
{[|V_{L33}|^2 \, (|(\widetilde{m}_{\nu}^{-1})_{23}|/|(\widetilde{m}_{\nu}^{-1})_{33}|)
\sin \a_{L}^{\tau A} + |V_{L33}|\,|V_{L23}|\, \sin \a_{L}^{\tau B} ]
\over |(\widetilde{m}_{\nu}^{-1})_{33}|^{2} + |(\widetilde{m}_{\nu}^{-1})_{23}|^{2}} \,  ,
\ee
where
\bea
\a_L^{\tau A} & = & 
{\rm Arg}\left[\widetilde{m}_{\nu 11}\right]  - 2\,{\rm Arg}[(\widetilde{m}^{-1}_{\nu})_{23}] 
- \pi -2\,(\rho+\s) -2\,(\rho_L+\s_L)  \,  ,
\\
\a_L^{\tau B} & = & 
{\rm Arg}\left[\widetilde{m}_{\nu 11} \right]  - \,{\rm Arg}[(\widetilde{m}^{-1}_{\nu})_{23}] 
- {\rm Arg}[(\widetilde{m}^{-1}_{\nu})_{33}]   -2\,(\rho+\s) -2\,(\rho_L+\s_L)  \,   .
\eea
Analogously, one finds an analytic expression for the muon $C\!P$ asymmetry given by \cite{full}
\bea
\ve_{2\mu} \simeq \ve_{2\mu}^{V_L}  & = &  {3 \,m^2_{D2} \over 16\, \pi \, v^2}\,
{|(\widetilde{m}_{\nu})_{11}| \over m_1 \, m_2 \, m_3}\,   \\ \nonumber
& \times  &  {|(\widetilde{m}_{\nu}^{-1})_{23}| \over |(\widetilde{m}_{\nu}^{-1})_{33}|} \, 
 {|V_{L22}| \, |V_{L 32}| \, \sin\a_{L}^{\m A} + 
 |V_{L 32}|^2 \, (|(\widetilde{m}_{\nu}^{-1})_{23}| / |(\widetilde{m}_{\nu}^{-1})_{33}| )\sin\a_{L}^{\m B}\over |(\widetilde{m}_{\nu}^{-1})_{33}|^{2} + |(\widetilde{m}_{\nu}^{-1})_{23}|^{2}}  \,   ,
\eea 
where 
\bea
\a_L^{\m A} & = & 
{\rm Arg}\left[\widetilde{m}_{\nu 11}\right]  - {\rm Arg}[(\widetilde{m}^{-1}_{\nu})_{23}] 
- {\rm Arg}[(\widetilde{m}^{-1}_{\nu})_{33}]  -2\,(\rho+\s) -2\,(\rho_L+\s_L)  \,  ,
\\
\a_L^{\m B} & = & 
{\rm Arg}\left[\widetilde{m}_{\nu 11} \right]  - 2\,{\rm Arg}[(\widetilde{m}^{-1}_{\nu})_{23}] 
- \pi -2\,(\rho+\s) -2\,(\rho_L+\s_L)  \,   .
\eea
Since there are no electron dominated solutions,  we do not give here the analytic expression for 
the electron $C\!P$ asymmetry but this can be found in \cite{full}.\footnote{However, notice that electron dominated
solutions are found in a supersymmetric framework \cite{susy}.}

In the expressions (\ref{twofl}) we have also introduced 
the {\em flavoured decay parameters} $K_{I\a}$ defined as 
\be
K_{I\a} \equiv {\G_{I\a}+\overline{\G}_{I\a}\over H(T=M_I)}= 
{|m_{D\a I}|^2 \over M_I \, m_{\star}} \,  ,
\ee
 where 
$m_{\star}\equiv 16\,\pi^{5/2}\, \sqrt{g^{SM}_{\star}}/(3 \sqrt{5})\,(v^2 / M_{\rm Pl}) \simeq 1.07 \,{\rm meV}$
is the equilibrium neutrino mass  and 
$H(T)=\sqrt{g^{SM}_{\star}\,8\,\pi^3/90}\,T^2/M_{\rm P}$ is the expansion rate.

It is easy to obtain the following expression for the flavoured decay parameters in 
the bi-unitary parameterisation (see Eq.~(\ref{biunitary})) \cite{full}
\be\label{KialVL}
K_{I\a} = {\sum_{k,l} \, 
m_{Dk}\, m_{Dl} \,V_{L k\a} \, V_{L l\a}^{\star} \, U^{\star}_{R kI} \, U_{R lI} 
\over M_I \, m_{\star}}\,  .
\ee
 
Finally, for the efficiency factors at the production $\k(K_{2\a})$
we can use the standard simple analytic expression valid for initial thermal abundance \cite{predictions}
\be\label{kappa}
\k(K_{2\a}) = 
{2\over z_B(K_{2\a})\,K_{2\a}}
\left(1-e^{-{K_{2\a}\,z_B(K_{2\a})\over 2}}\right) \,  , \;\; z_B(K_{2\a}) \simeq 
2+4\,K_{2\a}^{0.13}\,e^{-{2.5\over K_{2\a}}} \,   .
\ee
Notice, however,  that since  all solutions are characterised by strong wash-out at the production
(either $K_{2\t} \gg 1$ or $K_{2 \tau_2^{\bot}} \gg 1$ respectively for tauon and muon-dominated solutions),
the final asymmetry does not depend on the initial $N_2$ abundance anyway.\footnote{For $\tau_B$ solutions,
that we introduce in the next section, one can find values of $K_{2\tau}$ as low as $K_{2\tau}\simeq 2$ 
but only for values $m_1 \gtrsim m_{\rm atm}$, now disfavoured by the cosmological observations.} 

In this way we have now all the analytical expressions
needed to calculate the final asymmetry $N_{B-L}^{\rm f}$ in a fast way 
and from this, using Eq.~(\ref{etaBlep}), the baryon-to-photon ratio $\eta_B^{\rm lep}$
predicted by leptogenesis as a function of $\a_2$, the nine low energy neutrino parameters in $m_\nu$
and the nine parameters in $V_L$.


\section{Scatter plots: 3-dim projections}

The baryon-to-photon  ratio predicted by leptogenesis, that can be calculated with the analytic expression
$\eta_{B}^{\rm lep}(\a_2,m_\nu,V_L)$ given in the previous section,
has to be compared with the experimental  value from cosmological observations \cite{planck18}
\be\label{etaBexp}
\eta_{B}^{\rm exp} = (6.12 \pm 0.04) \times 10^{-10} \,   .
\ee
If one approximates $V_L \simeq I$ and assumes $\a_2$ values  lower than a certain maximum allowed value, 
then the successful leptogenesis condition 
$\eta_{B}^{\rm lep}(\a_2,m_\nu,V_L) = \eta_{B}^{\rm exp}$ defines an hypersurface in the space of the 
nine low energy neutrino parameters. When the dependence on the
parameters in the $V_L$ is taken into account, the hypersurface becomes a layer with some thickness determining an allowed region
in the space of parameters.
Since $SO(10)$-inspired conditions impose stringent upper bounds on the three mixing angles $\theta_{ij}^L$, 
this thickness is sufficiently moderate that one still obtains experimental, 
partly testable, predictions \cite{riotto1,riotto2}.

The determination of this allowed region can be done numerically with scatter plots \cite{riotto1,riotto2,decrypting,full}. 
So far the resulting constraints have been shown projecting on different planes, in particular $\theta_{23}$ versus 
$m_1$ and $\delta$ versus $m_1$. However, these 2-dim projections can hide the full higher dimensional structure 
of the constraints and, therefore, the predictive power of the scenario. Without imposing any 
experimental information on $\delta$, allowing uniformly any value in $[0,2\pi]$, 
there would be no  loss of predictive power in neglecting the dependence on $\delta$.
However, the experimental data now favour a certain range of values for $\delta$  
excluding at $3\s$ quite a large interval (see Eq.~(\ref{deltarange})).

For this reason, in the left panel of Fig.~1, we now show 3-dim projections of the scatter plots in the space 
$(\delta,\theta_{23},m_1)$ for $\a_2 =5$ and $0 \leq \theta_{ij}^L \leq \theta_{ij}^{\rm CKM}$. 
 It can be seen that for sufficiently large values of $\theta_{23}$,
and in particular for $\theta_{23}$ in the second octant, one has two different regions corresponding to two disconnected
ranges of values for $m_1$: 
one at high values, approximately for  $34\,{\rm meV} \lesssim m_1 \lesssim 100 \, {\rm meV}$, 
and one at low values, approximately for $1\,{\rm meV} \lesssim m_1 \lesssim 10 \, {\rm meV}$,
with the exact limits depending on the values of $\delta$ and $\theta_{23}$ and in general such that
the ranges reduce for increasing values of $\theta_{23}$.
\begin{figure}
\begin{center}
        \psfig{file=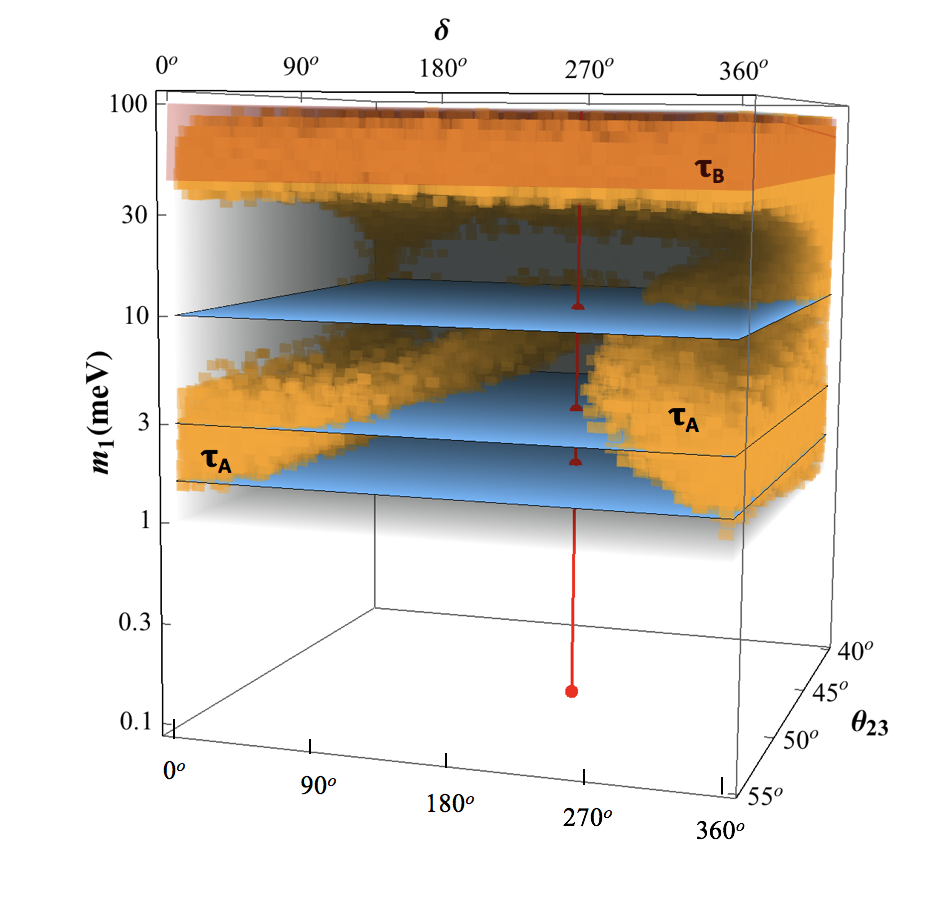,height=70mm,width=70mm}  \hspace{5mm}
        \psfig{file=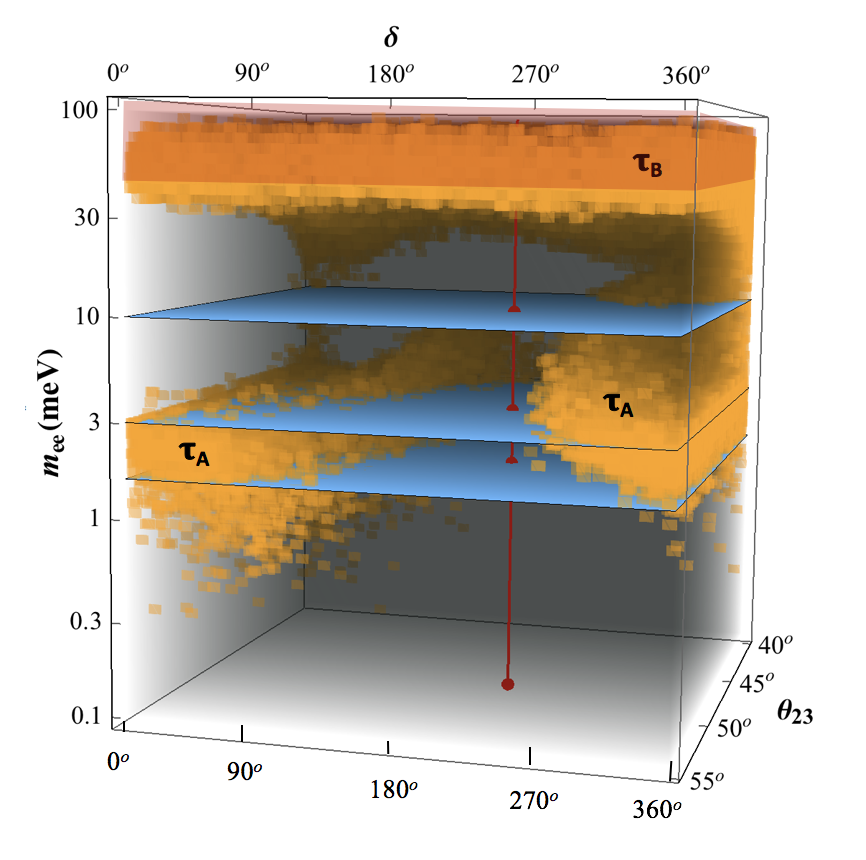,height=70mm,width=70mm} 
\end{center}
\caption{Scatter plots in the space $(\d,\theta_{23},m_1)$ (left panel)
and $(\d,\theta_{23},m_{ee})$ (right panel) for $\a_2 = 5$ and $0 \leq \theta_{ij}^L \leq \theta_{ij}^{\rm CKM}$.
The red vertical axis indicates best fit values for $\delta$ and $\theta_{23}$. The red shaded regions correspond to
$m_1 \gtrsim 43\,{\rm meV}$, currently disfavoured by cosmological observations. The three blue planes, at $m_1 =1.5\,{\rm meV}, 3\,{\rm meV}$ and $10\,{\rm meV}$ help to visualise the ranges of $m_1$ values allowed by $\tau_A$ solutions in the
$(\delta,\theta_{23})$ plane.}
\label{fig:3DSP}
\end{figure}

These two regions correspond to two different
types of analytical solutions that were clearly identified 
and characterised in \cite{riotto2,decrypting}.
They both correspond to tauon dominated solutions but
those at low $m_1$ values, the $\tau_A$ solutions, are approximately characterised by 
$\rho \simeq n\,\pi/2$ and $K_{2\tau} \gg 1$, while those at high $m_1$ values,
the $\tau_B$ solutions, are characterised by $\rho \simeq n\,\pi$ and $K_{2\tau} \simeq 1$. 
Therefore, an important difference is that in the case of $\tau_A$ solutions (low $m_1$ values)
the wash-out at the production is necessarily very strong ($K_{2\tau} \gtrsim 10$) and the final asymmetry is independent
of the initial $N_2$ abundance, while in the case of $\tau_B$ solutions  
the wash-out at the production is mild ($K_{2\tau} \gtrsim 2$) and for this reason 
there can be some slight dependence on the initial $N_2$ abundance.\footnote{Strong thermal leptogenesis can be realised
only for $\tau_A$ solutions \cite{decrypting}. For increasing values of the  initial pre-existing asymmetry to be washed-out,
one needs increasing value of $m_1$ in order to have higher values of $K_{1e} \propto m_{ee} \simeq m_1$ 
and this, in turn, requires lower and lower values of $\theta_{23}$. In this way one finds  (more stringent) 
lower bounds on $m_1$ and $m_{ee}$  and an upper bound on 
$\theta_{23}$ depending on the initial pre-existing asymmetry to be washed-out \cite{decrypting}. 
For example for a $10^{-3}$ value of the initial pre-existing asymmetry and for $\a_2 =5$ one finds
$m_1, m_{ee}\gtrsim 8\,{\rm meV}$ and $\theta_{23} \lesssim 45.75^\circ$ incompatible with $\theta_{23}$
in the second octant \cite{chianese}.
However, for higher values of $\a_2$, the upper bound on $\theta_{23}$ gets relaxed and, in particular,
for $\a_2 =6$ one obtains $\theta_{23} \lesssim 54^\circ$ compatible with current experimental values \cite{chianese}.} 

An interesting feature is that while in the case of $\tau_B$ solutions, for  $m_1 \gtrsim 34\,{\rm meV}$,
all values of $\delta$ and $\theta_{23}$ are allowed,  in the case of $\tau_A$ solutions, for low $m_1$,  
not all values of $\delta$ and $\theta_{23}$ are allowed. In particular, there is 
a range of values of $\delta$, modulated by $\theta_{23}$, that is unaccessible to $\tau_A$ solutions.
In this way, if future cosmological observations will place an upper bound below $34\,{\rm meV}$, excluding
$\tau_B$ solutions, then,  for a given couple of values   $(\d,\theta_{23})$, there is quite  a narrow allowed range of $m_1$ values
(in Fig.~1 this range can be understood with the help of the three blue planes at $m_1 =1.5\,{\rm meV}$, $3\,{\rm meV}$
and $10\,{\rm meV}$).
 Considering that current best fit values 
 of $\delta$ and $\theta_{23}$ fall just in the region unaccessible
 to $\tau_A$ solutions (in Fig.~1 these best fit values are indicated by the red vertical axis), one has a very interesting situation: 
 {\em if the experimental errors on  $\delta$ and $\theta_{23}$ will sufficiently shrink around current best fit values, 
 then either absolute neutrino mass scale experiments will find a positive signal 
 or $SO(10)$-inspired leptogenesis with $\a_2 \lesssim 5$ will be ruled out}.   This shows that 
 in the next years low energy neutrino experiments will test $SO(10)$-inspired 
 leptogenesis in a very effective and interesting way.
 
Another interesting feature is that even outside the region in the plane $(\theta_{23},\delta)$ for which 
$m_1 \gtrsim 34\,{\rm meV}$ ($\tau_B$ solutions),
there is still  a lower bound $m_1 > m_1^{\rm min}(\d, \theta_{23}) \sim {\rm meV}$, holding for $\tau_A$ solutions,
with the exact value depending on $\delta$ and $\theta_{23}$ as we will discuss in detail in the next section. 
Even though this lower bound is much more relaxed,  it might be still tested in future and it is in any case important 
when considering specific $SO(10)$-inspired models.

In the right panel of Fig.~1 we also show a scatter plot in the space $(\delta,\theta_{23},m_{ee})$.
Interestingly, it can be noticed that the region in the plane $(\delta,\theta_{23})$ where $m_1 \gtrsim 34\,{\rm meV}$
translates into a slightly more relaxed lower bound $m_{ee} \gtrsim 31\,{\rm meV}$ that will be fully tested by next generation
$0\nu\beta\beta$ experiments. Outside this region the lower bound on $m_{ee}$ can relax to values that are too small to give a signal
but there are still regions where the lower bound is stringent enough that might be testable by next generation experiments.
We discuss this in more detail in the next section but it is interesting that if long baseline experiments
will confirm values of $\delta$ and $\theta_{23}$ not too different from current best fit values,
then both cosmological observations and $0\nu\b\b$ experiments should find a positive signal and measure the
absolute neutrino mass scale if the $SO(10)$-inspired leptogenesis scenario is correct, 
otherwise they will rule it out or place strong constraints on the parameters defining $SO(10)$-inspired
conditions (in particular on $\a_2$ and the angles in $V_L$ as we discuss in Section 5). 

Finally, notice that in Fig.~1 we indicated, with a red colour, the region in tension with the upper bound from cosmological
observations  $m_1 \lesssim 43\,{\rm meV} (95\% \, {\rm C.L.})$. Notice that within $SO(10)$-inspired leptogenesis this upper
bound also applies on $m_{ee}$, since at these high values of $m_1$ the ($\tau_B$) solutions satisfy $m_{ee}\simeq m_1$,
as we show in Section 6, another, quite distinctive, feature.

\section{Lower bound on the absolute neutrino mass scale}    

The scatter plots in Fig.~1 clearly confirm the existence of the lower bounds on $m_1$ and $m_{ee}$.
The results in \cite{full} were already showing that the lower bound on $m_1$ was strongly modulated by $\delta$
and that for certain values this could become stringent enough to  be testable 
by absolute neutrino mass scale experiments.
They were also showing that the lower bound was becoming more stringent for large values of $\theta_{23}$.

We extracted the lower bounds on $m_1$ and $m_{ee}$ from the scatter plots in Fig.~1,
showing their simultaneous dependence both on $\delta$ and $\theta_{23}$. 
To this extent, we show in Fig.~2 isocontour lines
of the lower bound on $m_1$ (left panel) and $m_{ee}$ (right panel) in the plane $\delta$ versus $\theta_{23}$
for $\a_2 =5$ and $0 \leq \theta_{ij}^L \leq \theta_{ij}^{\rm CKM}$. The blue region corresponds to the case
when the lower bound is realised by $\tau_A$ solutions, while the orange region corresponds to the region where 
$\tau_A$ solutions are missing and the lower bound is determined by $\tau_B$ solutions. 
Since the two sets of solutions are disconnected at such high values of $\theta_{23}$,
there is a discontinuity in the value of the lower bounds.
 \begin{figure}
\begin{center}
        \psfig{file=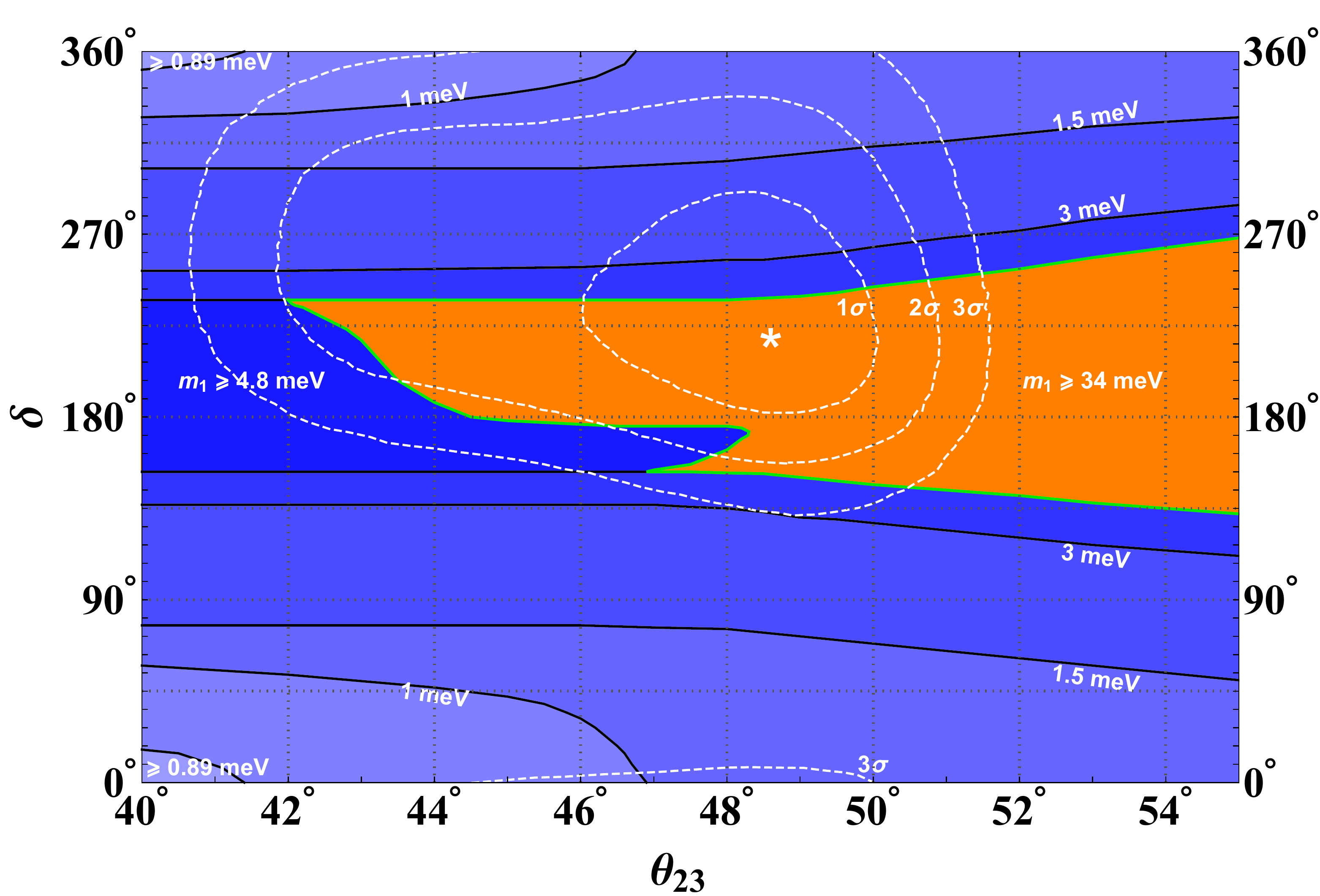,height=56mm,width=78mm}   \hspace{1mm}
        \psfig{file=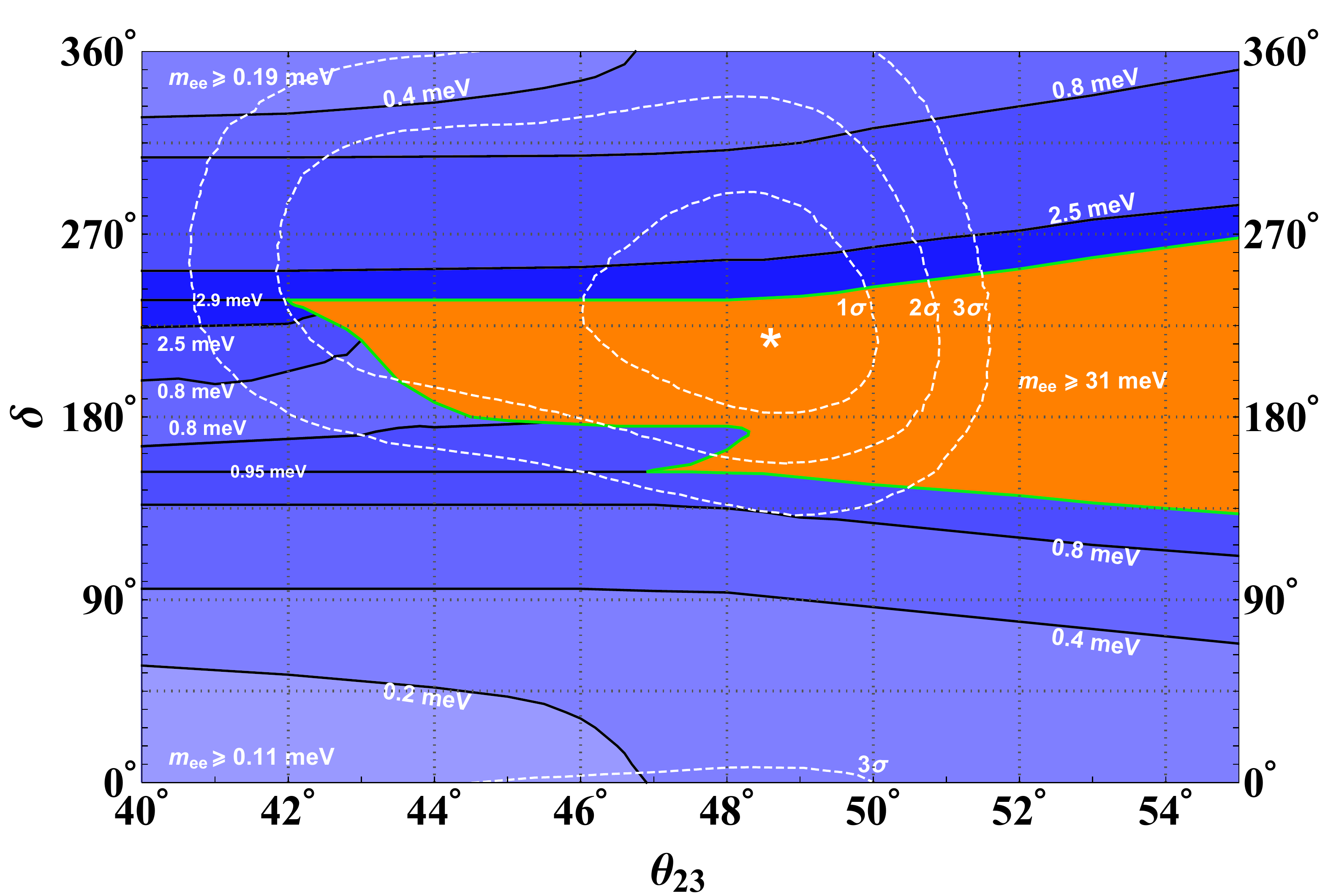,height=56mm,width=78mm} 
\end{center}
\caption{Isocontour lines for the lower bound on the lightest neutrino mass $m_1$ (left panel) and on the $0\nu\beta\beta$ effective neutrino mass $m_{ee}$ (right panel) for $\a_2 = 5$ in the plane $\delta$ versus $\theta_{23}$. The white dashed lines are the current experimentally favoured regions for $\delta$ and $\theta_{23}$ at the indicated C.L. from global analyses \cite{nufit} and the white stars indicate the best fit values. The blue area indicates
$\tau_A$ solutions, while the orange area indicates $\tau_B$ solutions.}
\label{fig:mass_regime}
\end{figure}
In the same figure we also superimpose white short-dashed lines showing the allowed regions found by global 
analyses at $1\sigma$, $2\sigma$ and $3\sigma$ C.L. \cite{nufit}. The white stars indicates the best fit 
values for $\d$ and $\theta_{23}$.  
It can be seen how current data favour $\theta_{23}$ in the second octant and $\delta$ in the third
quadrant. From Fig.~2 it is clear how  this experimentally favoured region strongly overlaps with the orange region, 
for $\delta$  in the range $150^\circ$--$240^\circ$ at $\theta_{23}= 48.6^\circ$ (best fit value), 
where the lower bounds get much more stringent and precisely 
\be\label{testregion}
m_1 \gtrsim 34\,{\rm meV} \hspace{10mm}\mbox{\rm and}\hspace{10mm} m_{ee} \gtrsim 31\,{\rm meV}  \,   .
\ee
These lower bounds are sufficiently stringent that  will be tested during next years by absolute neutrino
mass scale experiments. The first lower bound on $m_1$ corresponds to $\sum_i m_i \gtrsim 125\,{\rm meV}$
and, compared to the existing upper bound $\sum_i m_i < 146 \,{\rm meV}$ ($95\%$ C.L.) \cite{hannestad},
it is clear that there is already some tension. The second lower bound on $m_{ee}$ will be tested by next generation
$0\nu\b\b$ experiments. For example, the KamLAND2-ZEN experiment should reach a sensitivity of
$m_{ee} \simeq 20\,{\rm meV}$ \cite{kamland2zen} that would certainly fully test this high $m_1$ value
region ($\tau_B$ solutions). 

Outside the fully testable region (\ref{testregion}) the lower bound is of course much less stringent.
One can indeed only have $\tau_A$ solutions that are realised for values  in the range $m_1 \sim (1$--$10){\rm meV}$. 
The lowest bound is obtained for $\delta = n\,\pi$, confirming the results found in \cite{full} and in our case,
for $\theta_{23} \geq 40^{\circ}$, we find $m_1 \gtrsim 0.8\,{\rm meV}$. However, values $\delta =n\,\pi$
with $\theta_{23} \lesssim 43^{\circ}$ are currently excluded by global analyses at more than $3\sigma$.

Within the experimentally $2\s$ allowed region  we find $m_1 \gtrsim 1.2\,{\rm meV}$, 
that would correspond to $\sum_i m_i \gtrsim 59.8\,{\rm meV}$, with a deviation from the
hierarchical limit, where $\sum_i m_i = 58.5\,{\rm meV}$, of just $\d(\sum_i m_i) \gtrsim 1.3 \,{\rm meV}$.
Unfortunately such a small deviation is far beyond both current and planned cosmological observations 
(see \cite{lattanzi} for a review). However, if future long baseline measurements should shrink the allowed region in $(\theta_{23},\d)$
around current best fit values, then the deviation from the hierarchical limit could become
detectable.  It is then crucial whether future neutrino oscillation experiments will be able to measure
$\delta$ and $\theta_{23}$ precisely (and of course accurately) enough, in particular at the level 
to establish whether $\theta_{23}$ and $\d$ fall inside the region where the stringent lower bound 
in Eq.~(\ref{testregion}) hold (the region associated to $\tau_B$ solutions). 
For example, if errors should shrink around current best fit values,
a determination of $\delta $ with a $\D\d \simeq 5^\circ$ error and a determination of $\theta_{23}$
with a $\D\theta_{23}\simeq 0.5^{\circ}$ error would confirm the lower bounds (\ref{testregion}) at $3\s$ C.L..
Interestingly, these are precisions that will be reached combining results from next generation long 
baseline experiments DUNE and T2HK \cite{Ballett:2016daj}. In this case absolute neutrino mass scale
experiments should be able to measure both $m_1$ and $m_{ee}$ and to verify the prediction $m_{ee} \simeq m_1$
(see Section 6 for more details).

\section{Dependence of the lower bound on $SO(10)$-inspired conditions}          

In this section we discuss the dependence of the results on the definition of $SO(10)$-inspired conditions 
given in Section 2. From the calculation  of the asymmetry we have seen that this depends strongly on $\a_2$,
while it is independent of $\a_1$ and $\a_3$,
as far as of course one considers values of $\a_1$ and $\a_3$ for which the $N_2$ leptogenesis scenario and 
Eq.~(\ref{twofl}) is applicable.\footnote{For example, as we mentioned, 
if $\a_1 \lesssim 0.1$ so that $M_1 \lesssim T_{\rm sph}^{\rm out}\sim 100\,{\rm GeV}$, then there would be no
wash-out from $N_1$ inverse processes}  
In this case one simply has $N_{B-L}^{\rm f} \propto \a_2^2$ since 
$\ve_{2\a}\propto \a_2^2$ and  the two wash-out factors do not depend on $\a_2$ considering that all 
flavoured decay parameters are independent of all three $\a_i$. For this reason the lower bound on $m_1$,
as we will see in more detail in the next section, is simply $\propto \a_2^{\, -2}$, getting relaxed for increasing values of $\a_2$.

\begin{figure}
\begin{center}
      $\a_2 =4$ \\  \psfig{file=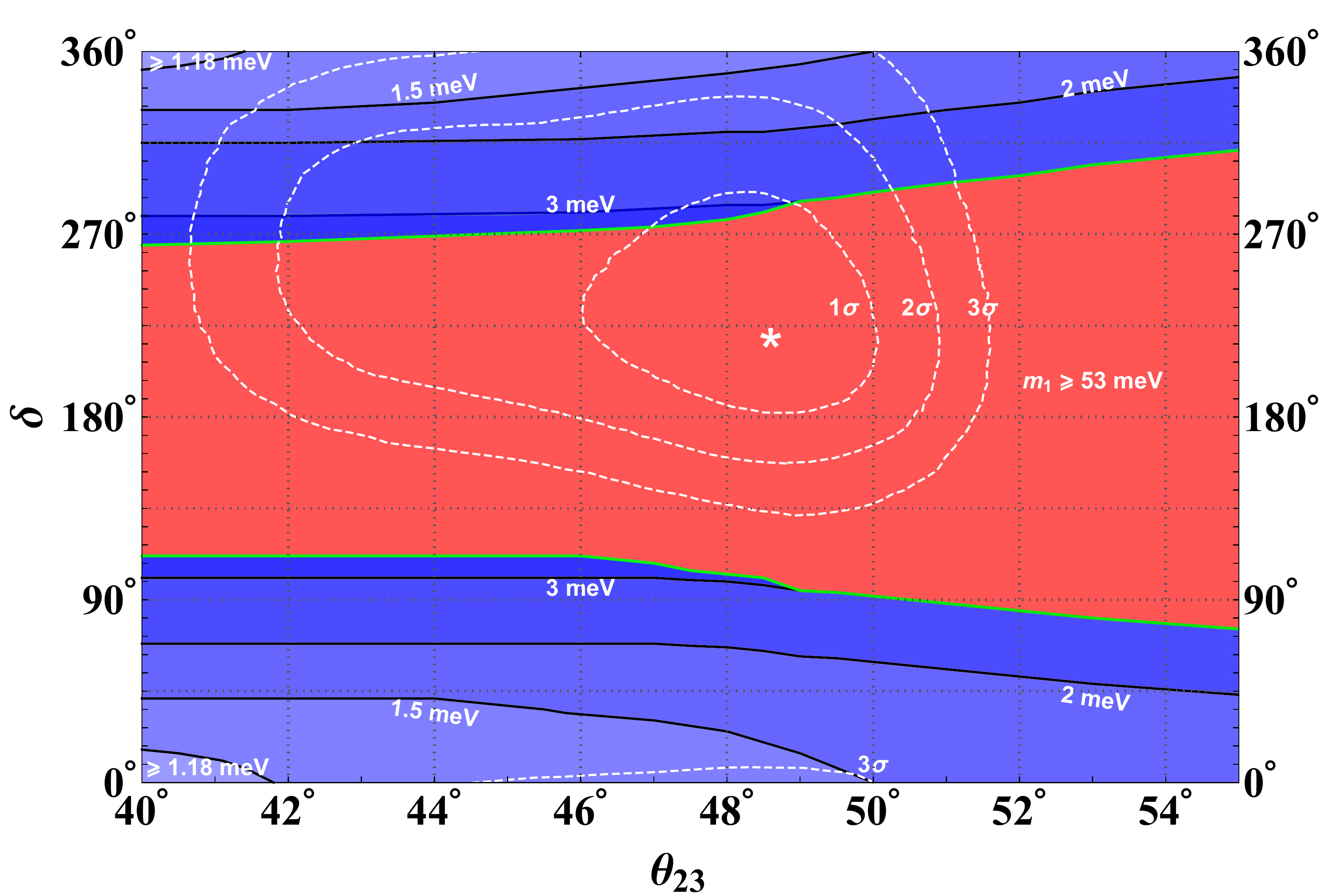,height=55mm,width=85mm}  \\ \vspace{5mm}
       $\a_2 =5$ \\ \psfig{file=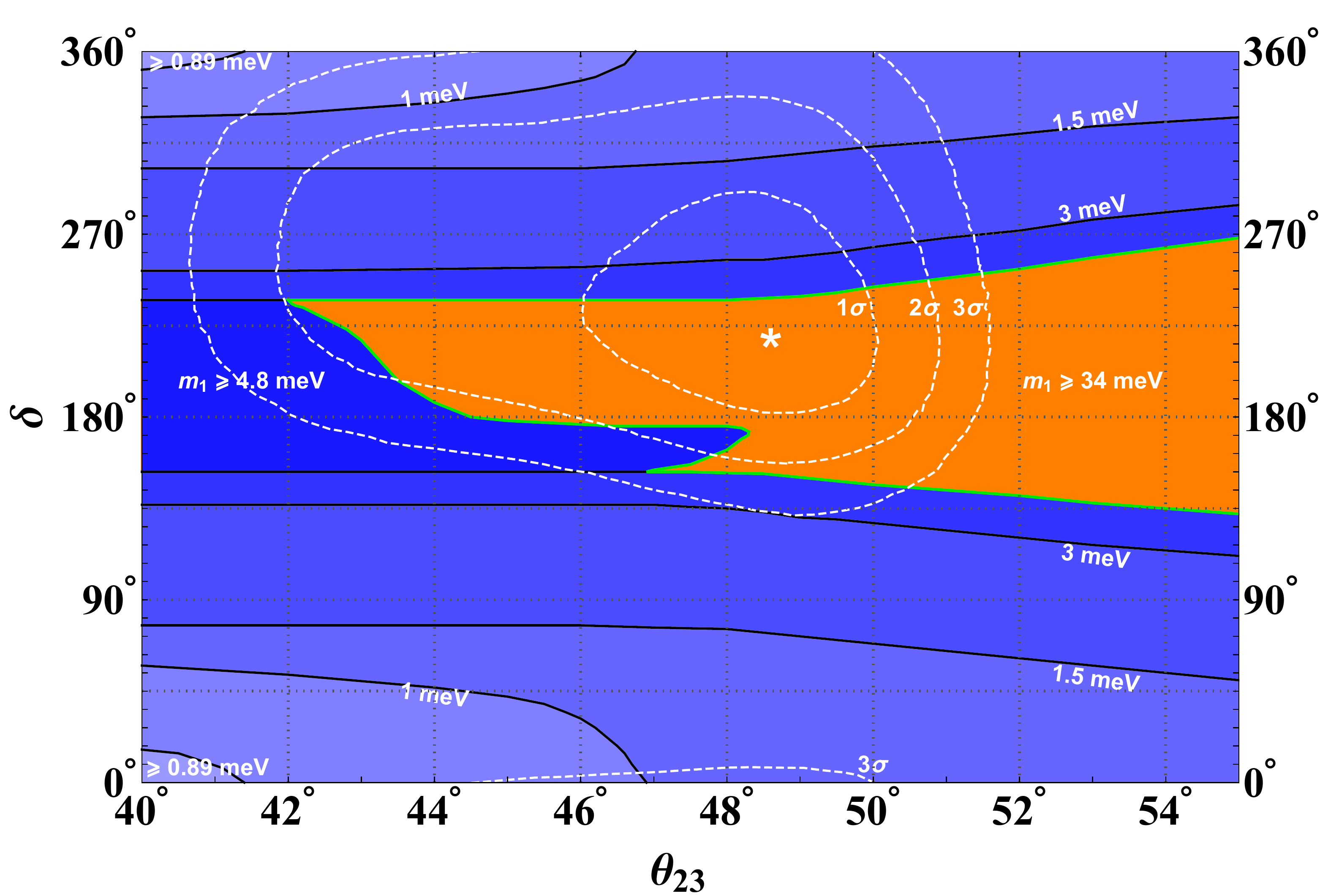,height=55mm,width=85mm}  \\ \vspace{5mm}
        $\a_2 =6$ \\ \psfig{file=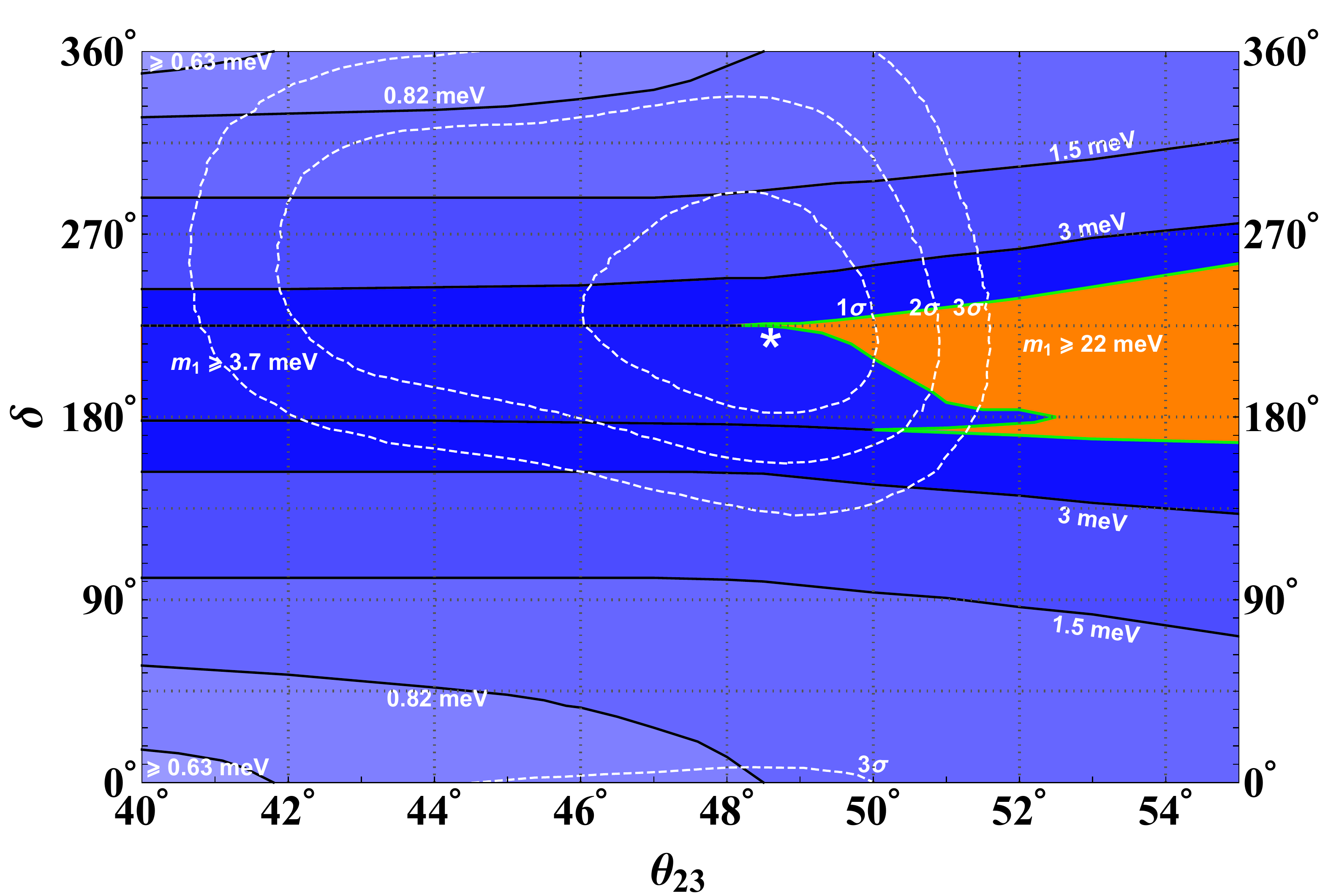,height=55mm,width=85mm}
\end{center}
\caption{Isocontour lines for the lower bound on the lightest neutrino mass in the plane $(\theta_{23},\delta)$ (same conventions as in Fig.~2)
for $\a_2 =4$ (top panel), $\a_2=5$ (central panel) and $\a_2 =6$ (bottom panel).}
\label{fig:alpha2m1}
\end{figure}

In Fig.~(\ref{fig:alpha2m1}) we show again, as in Fig.~3, the contour lines for the lower bound on $m_1$ in the plane $(\theta_{23},\delta)$
but this time for three different values of $\a_2$. The top panel is for $\a_2 =4$. One can see a very interesting result: the region filled by 
$\tau_A$ solutions  shrinks considerably so that the region not filled by $\tau_A$ solutions enlarges. At the same time $\tau_B$ solutions
also become disfavoured.  The reason is that for $\a_2 = 4$ the lower bound on $m_1$ for $\tau_B$ solutions becomes $m_1 \gtrsim 53\,{\rm meV}$, above the upper bound from cosmological observations $m_1 \lesssim 45 \,{\rm meV}$ ($95\%\,{\rm C.L.}$), having in mind, however,
that current tensions in the $\L$CDM model might be indicating some extension and some relaxation of the 
upper bound on neutrino masses cannot be excluded \cite{bariroma}. Therefore, this region is excluded at $95\%\,{\rm C.L.}$ by cosmological observations, or, in other words, it is only marginally allowed and, for this reason, we indicated it with red colour. Moreover since neutrino oscillation experiments favour values of  $\theta_{23}$ and $\delta$ to lie just in this region at $\sim 1\s$,
then one arrives to the conclusion that  current neutrino mixing data slightly disfavour values $\a_2 \lesssim 4$: this is
an interesting result showing well the interplay between absolute neutrino mass scale and neutrino mixing experiments 
in testing $SO(10)$-inspired leptogenesis, the main point of this paper. 
On the other hand, from the bottom panel one can see that for $\a_2 = 6$ the lower bound gets considerably relaxed in  the whole $(\theta_{23},\delta)$ plane.  In particular, $\tau_A$ solutions now exist for all values of $\delta$ even for $\theta_{23}$
in the second octant (more precisely, for $\theta_{23} \lesssim 48^\circ$)  and, marginally, even for best fit $(\theta_{23}, \delta)$ values. 
Moreover the lower bounds get considerably relaxed.

In Fig.~(\ref{fig:alpha2m1}) we show, for the same three values of $\alpha_2$, 
the contour lines for the lower bound on $m_{ee}$. The results are analogous to those for the lower bound on $m_1$.
In the case $\a_2 =4$, the region where there are no $\tau_A$ solutions is still indicated in red still to signal that this region 
is only marginally allowed by cosmological observations.  Notice that since at high values of $m_1$ one has $m_1 \simeq m_{ee}$,
we will show this result in Section 6, the lower bound $m_1 \gtrsim 53\,{\rm meV}$ translates also into an analogous lower bound
$m_{ee} \gtrsim 50\,{\rm meV}$. Interestingly, planned $0\nu\b\b$ experiments will be able to test this lower bound 
during next years.  
\begin{figure}
\begin{center} 
       $\a_2 =4$ \\ \psfig{file=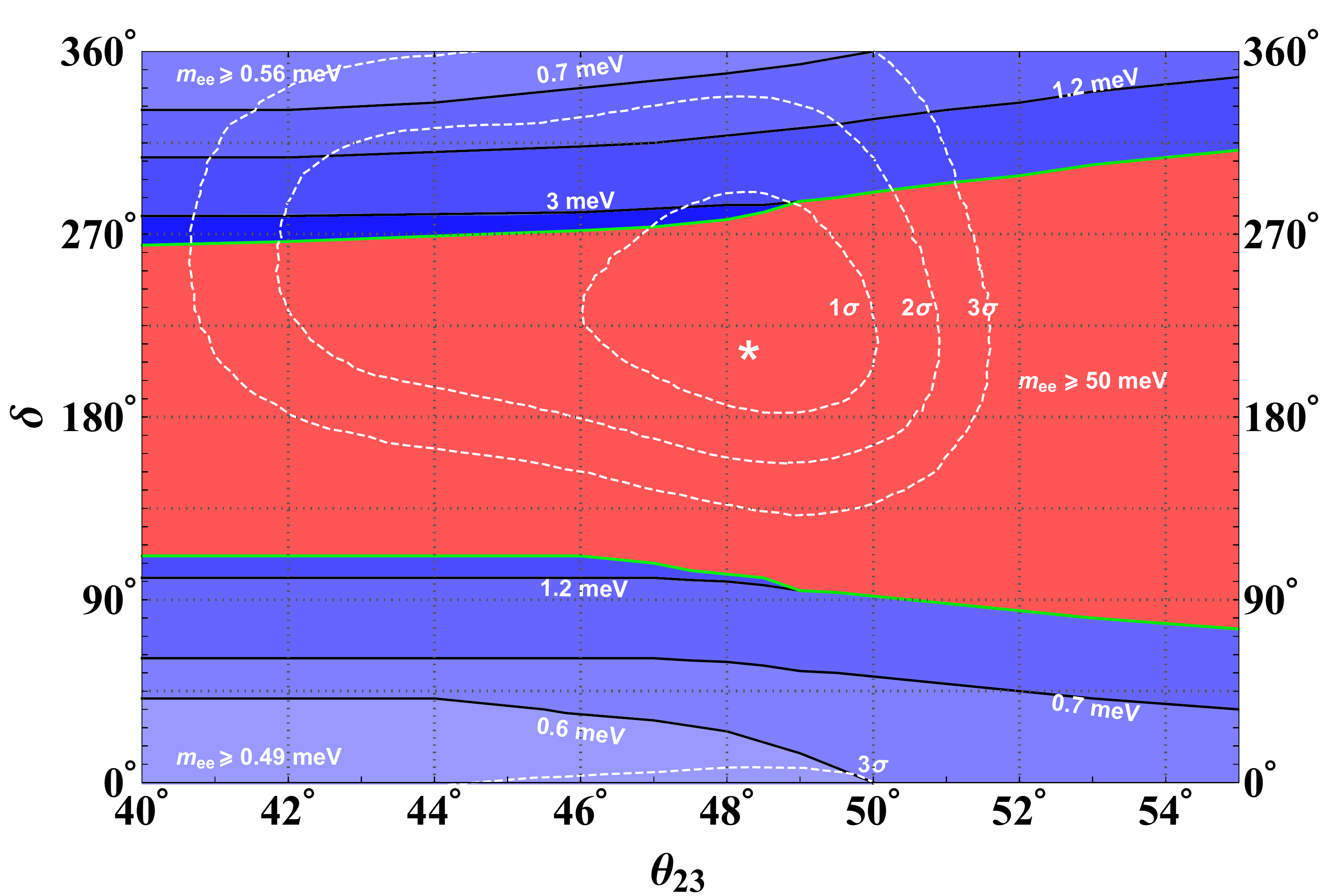,height=55mm,width=85mm}  \\ \vspace{5mm}
       $\a_2 =5$ \\ \psfig{file=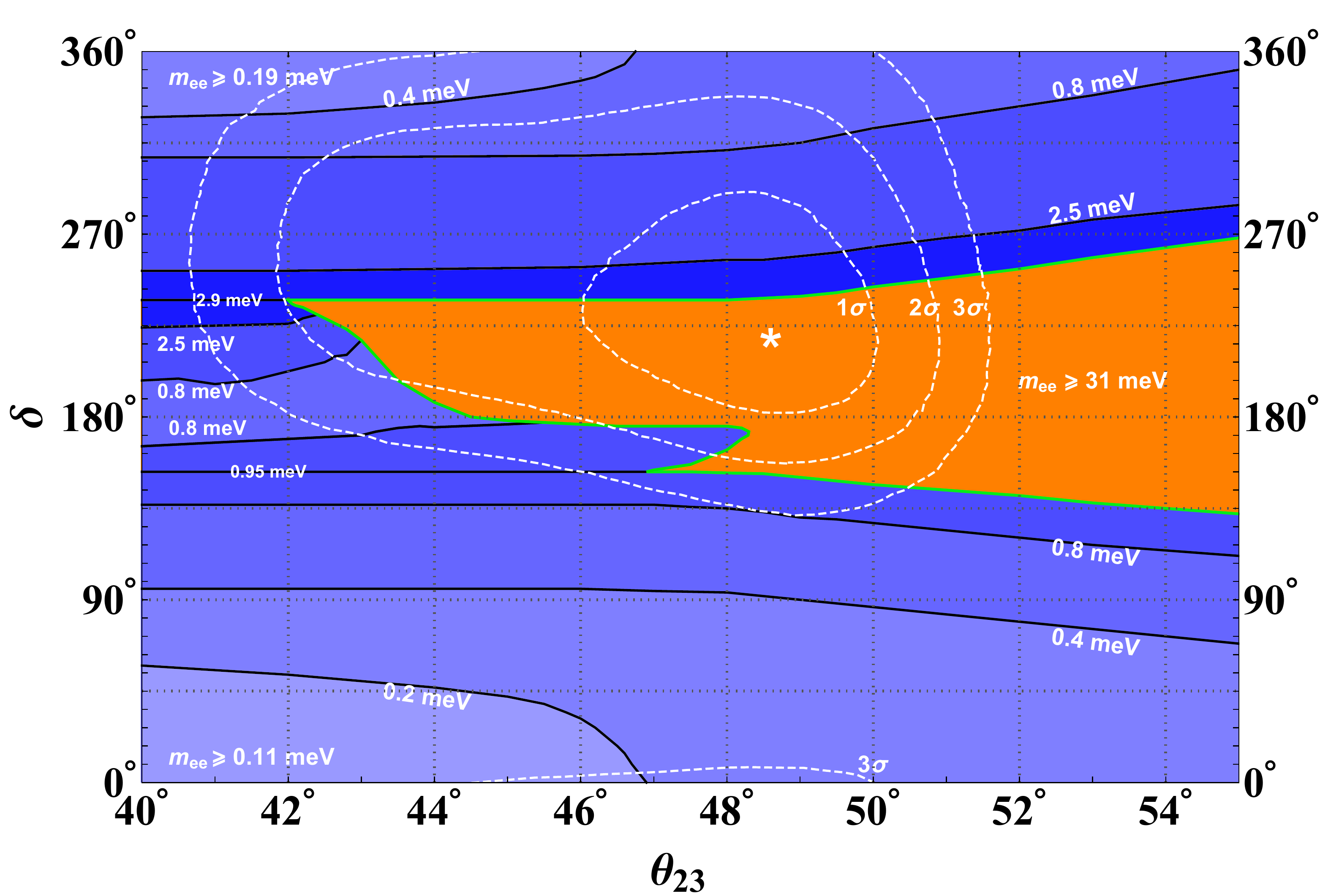,height=55mm,width=85mm}  \\ \vspace{5mm}
       $\a_2 =6$ \\\psfig{file=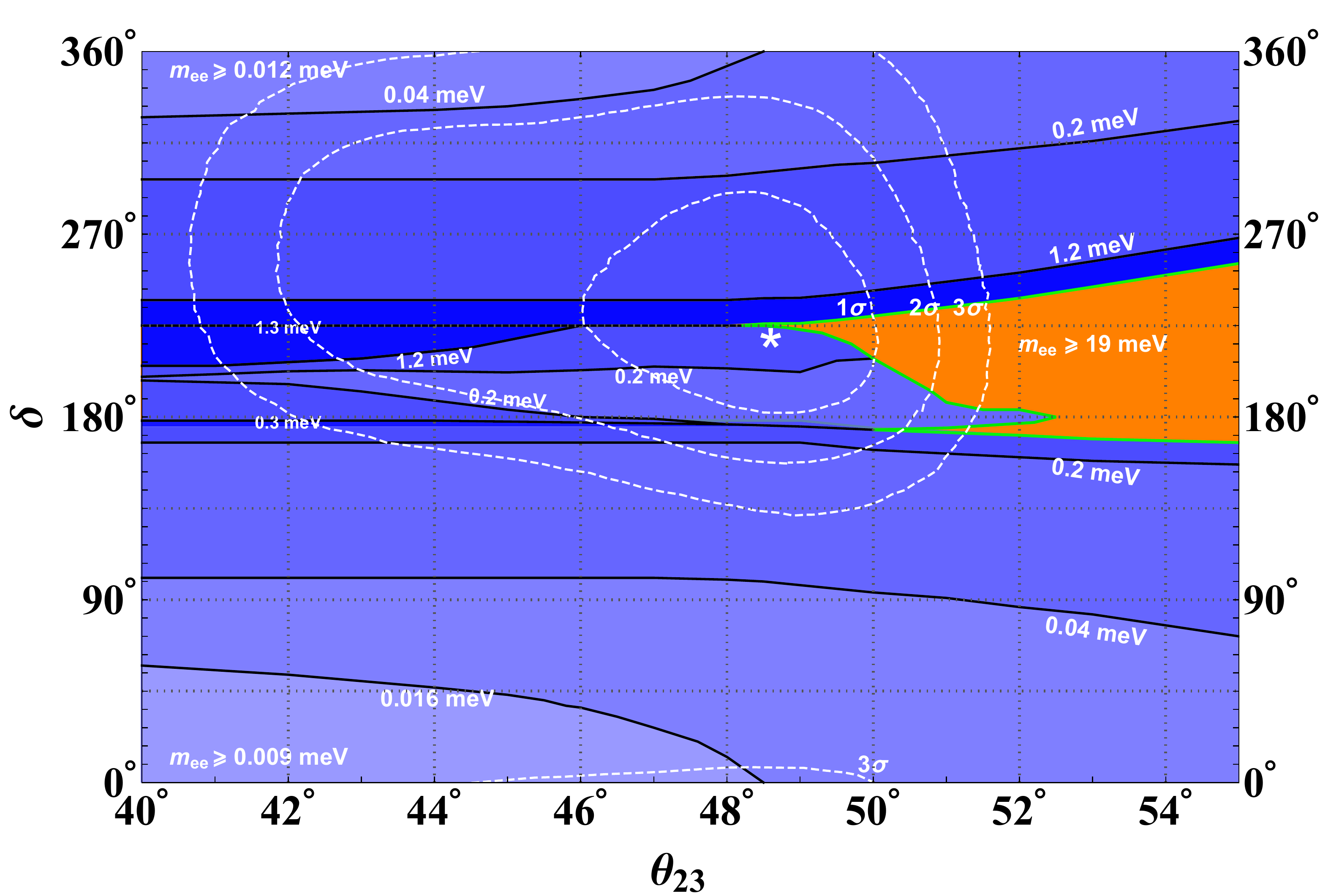,height=55mm,width=85mm}
\end{center}
\caption{Isocontour lines for the lower bound on $m_{ee}$ in the plane $(\theta_{23},\delta)$ (same conventions as in Fig.~2) 
for the three indicated values of $\a_2$.}
\label{fig:alpha2mee}
\end{figure}
On the other hand, for $\a_2 = 6$, the lower bound on $m_{ee}$, like that one for $m_1$, gets relaxed compared to the case $\a_2 =5$
and, as one can see from the bottom panel, one obtains $m_{ee} \gtrsim 19\,{\rm meV}$.

\begin{figure}
\begin{center}
      $0 \leq \theta_{ij}^L \leq \theta_{ij}^{CKM}$ \\  \psfig{file=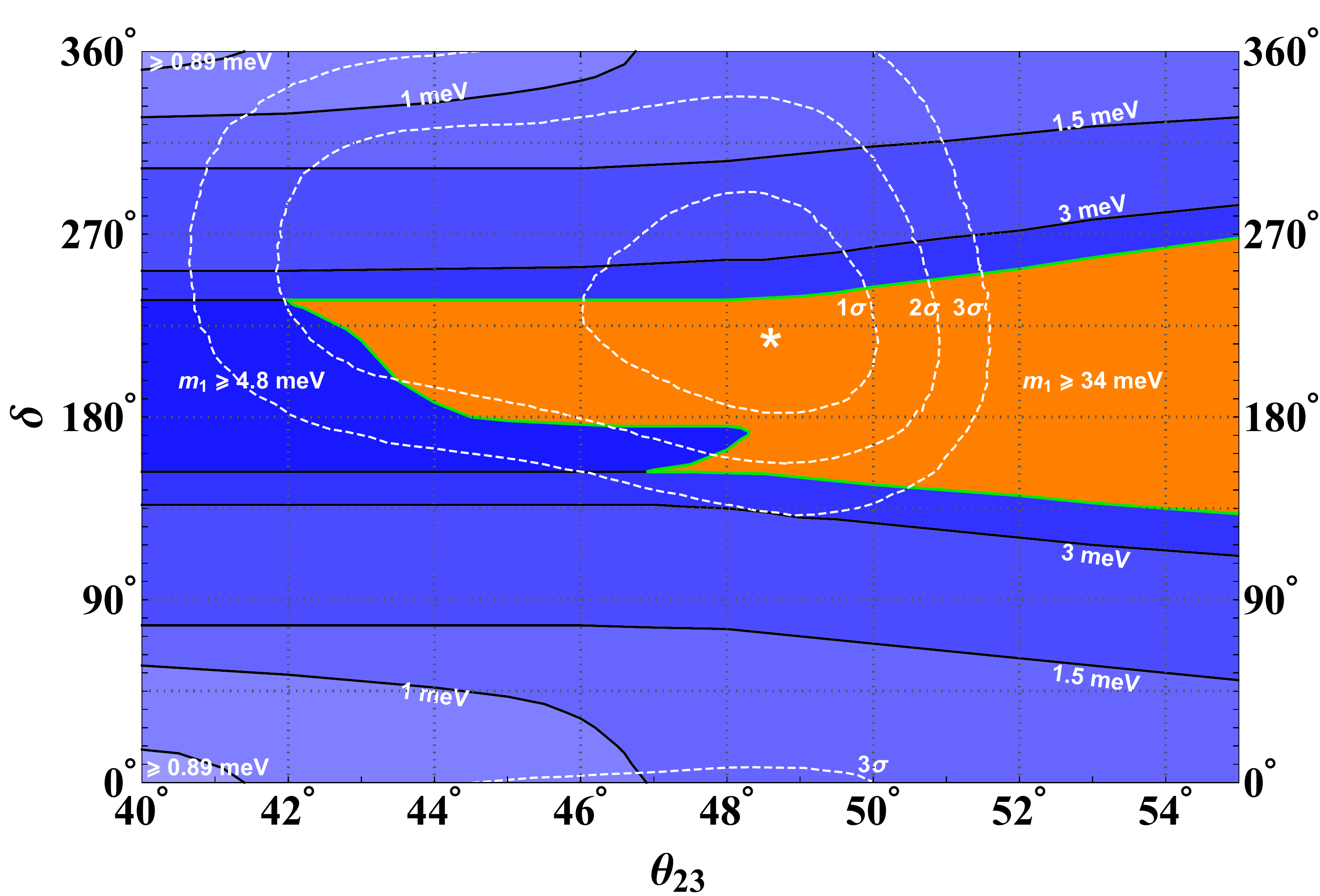,height=55mm,width=85mm}  \\ \vspace{5mm}
     $0 \leq \theta_{13}^L, \theta_{23}^L \leq 5^\circ$,  $0 \leq \theta_{12}^L \leq 13^\circ$ \\   \psfig{file=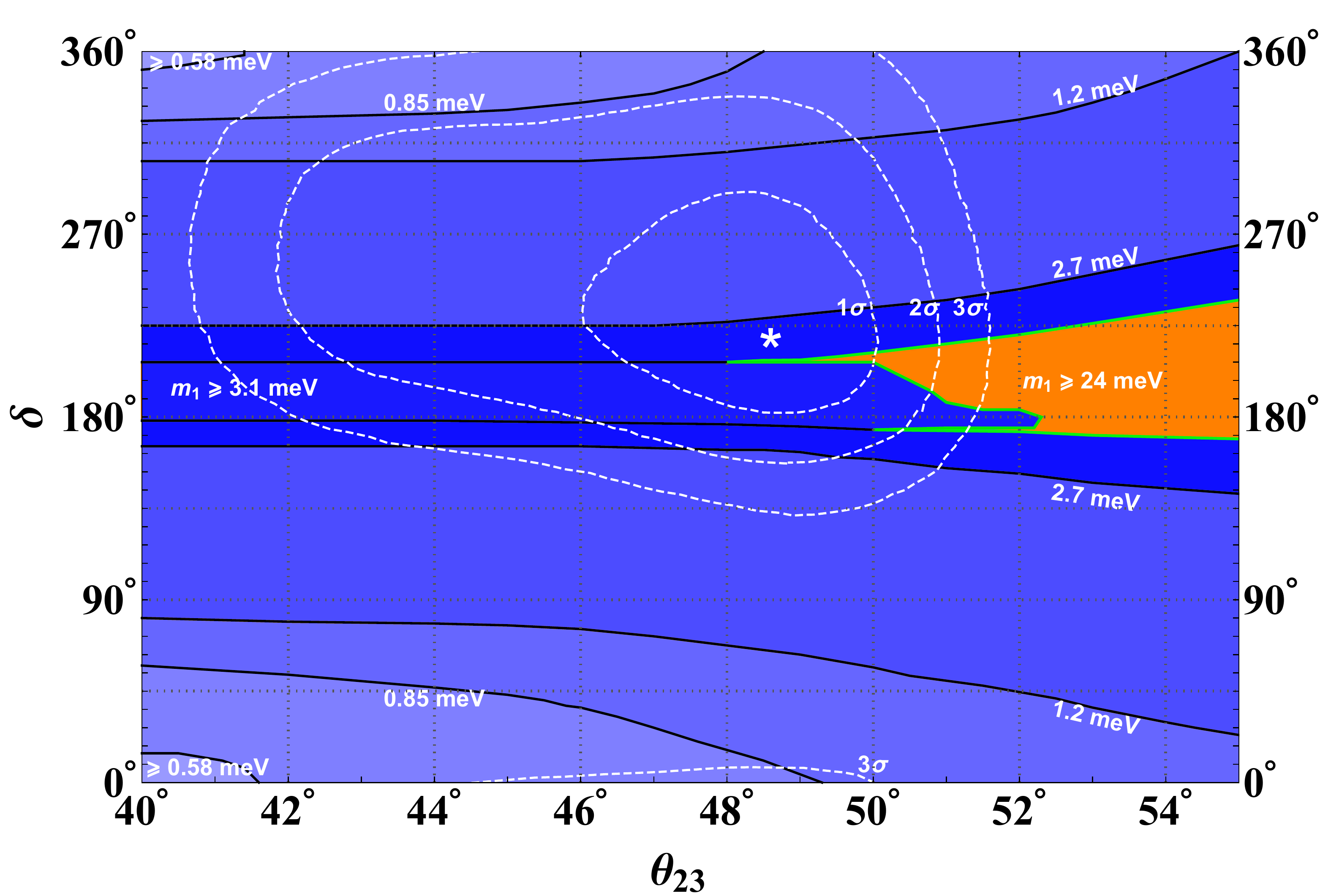,height=55mm,width=85mm}  \\ \vspace{5mm}
    $0\leq \theta_{ij}^L \leq 13^\circ$ \\    \psfig{file=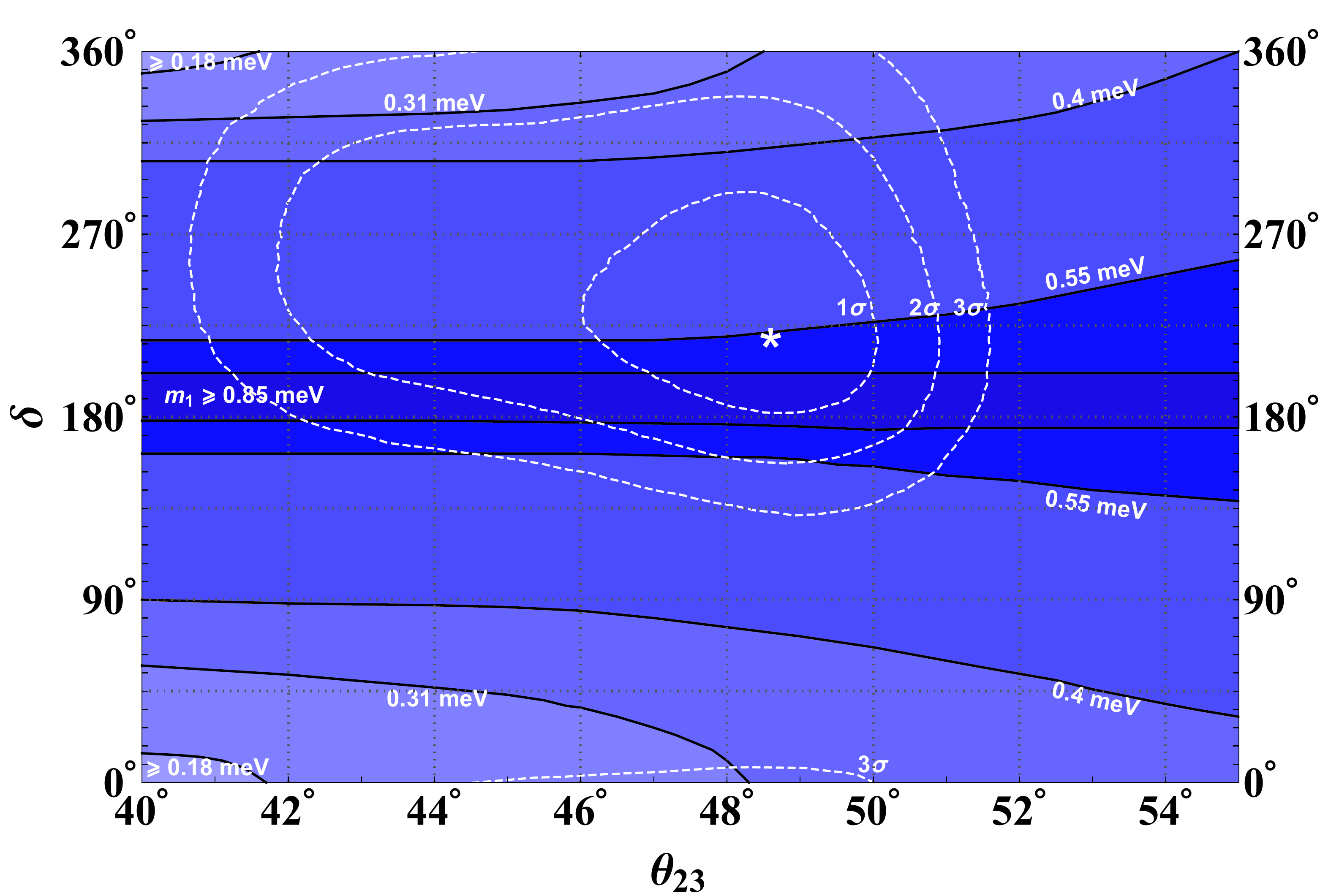,height=55mm,width=85mm}
\end{center}
\caption{Isocontour lines for the lower bound on $m_1$ in the plane $(\theta_{23},\delta)$ (same conventions as in Fig.~2) for
three different choices of the upper bounds on the three mixing angles $\theta_{ij}^L$: 
standard case $0 \leq \theta_{ij}^L \leq \theta_{ij}^{CKM}$ (upper panel);
$0 \leq \theta_{12}^L \leq \theta_{12}^{CKM} \simeq 13^\circ$ and $0 \leq \theta_{13}^L, \theta_{23}^L \leq 5^\circ$ (central panel);
$0 \leq \theta_{ij}^L \leq 13 ^ \circ$ (bottom panel).
}
\label{fig:VLm1}
\end{figure}

Together with $\a_2$, constraints on low energy neutrino parameters also depend on the three mixing angles 
$\theta_{12}^L, \theta_{13}^L, \theta_{23}^L$ in  the left-handed mixing matrix $V_L$. 
 Therefore, they depend on the precise definition of $SO(10)$-inspired conditions
in placing upper bounds on the $\theta_{ij}^L$. It was already discussed in detail in different papers \cite{riotto2,decrypting,full}
how the constraints change compared to the case $V_L = I$ when  small values of the mixing angles, at the level of the corresponding
angles in the CKM matrix, are turned on.  In Fig.~\ref{fig:VLm1} we show the results allowing a more extreme departure from $V_L=I$, especially aiming at understanding how large have to be the mixing angles to expect a drastic relaxation of the lower bound on the absolute neutrino mass scale. In the top panel we show again the results for $\a_2 =5$ and $0\leq \theta_{ij}^L < \theta_{ij}^{CKM}$.
In the central panel we still impose $\theta_{12}^L \leq \theta_{12}^{CKM} \simeq 13^\circ$ but this time we allow
$0 \leq \theta_{13}^L \leq 5^\circ$ and $0 \leq \theta_{23}^L \leq 5^\circ$. One can see how the blue region with 
$\tau_A$ solutions now fills almost all $(\theta_{23}, \delta)$ plane. In particular within the $1\s$ region 
allowed by neutrino oscillations experiments one now has $m_1 \gtrsim 2\,{\rm meV}$.
In the bottom panel we show the results for an even more drastic relaxation of $SO(10)$-inspired conditions,
allowing all three angles to vary within the range $0\leq \theta_{ij}^L \leq 13^\circ$. One can see how 
this time the $\tau_A$ solutions are allowed for all points in the $(\theta_{23},\delta)$ plane. However, despite this
drastic relaxation one can notice that there is still an absolute lower bound $m_1 \gtrsim 0.2\,{\rm meV}$
and within the $1\s$ region favoured by neutrino oscillation experiments one has $m_1 \gtrsim 0.5 \,{\rm meV}$.
\begin{figure}
\begin{center}
    $0 \leq \theta_{ij}^L \leq \theta_{ij}^{CKM}$  \\  \psfig{file=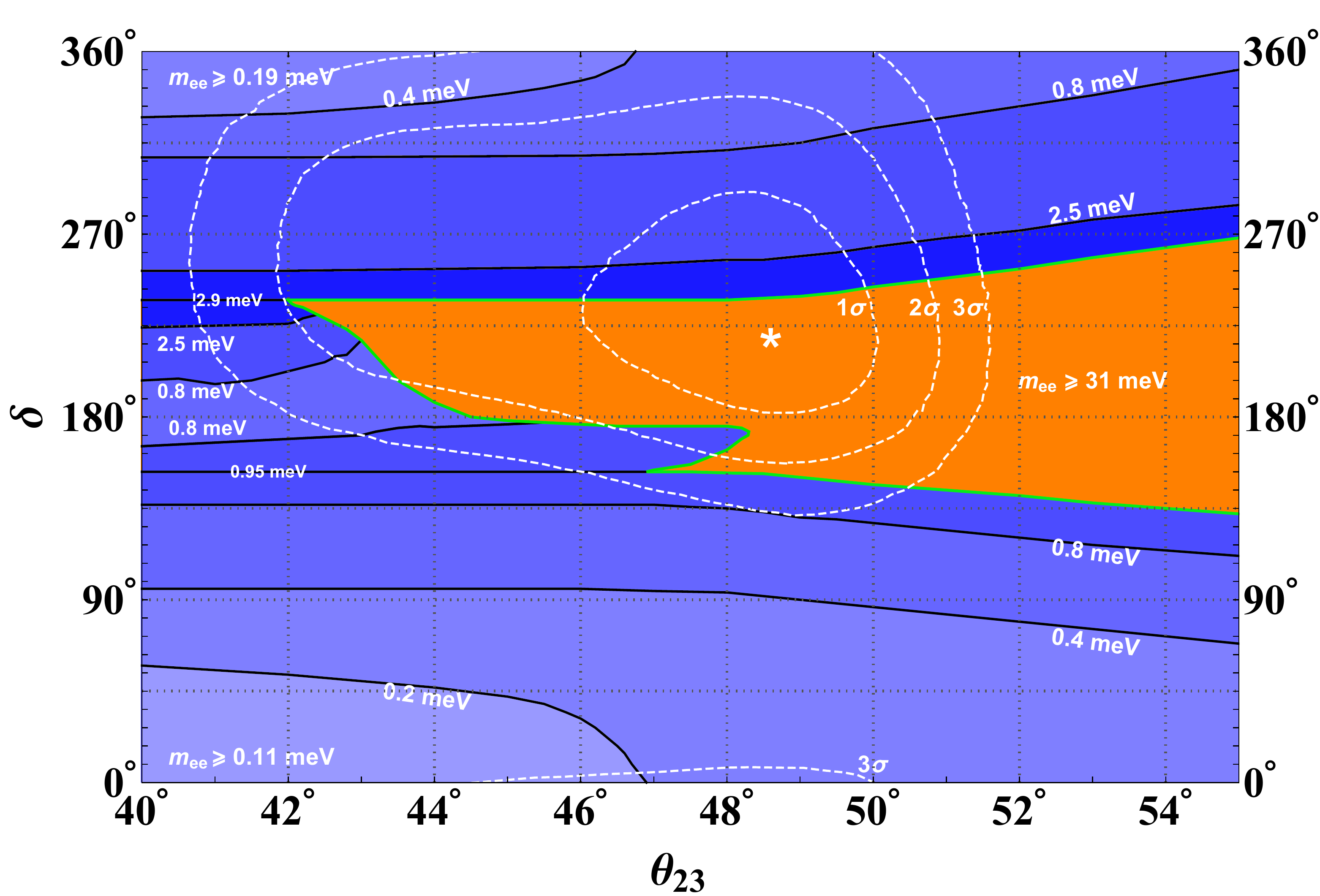,height=55mm,width=85mm}  \\ \vspace{5mm}
 $0 \leq \theta_{13}^L, \theta_{23}^L \leq 5^\circ$,  $0 \leq \theta_{12}^L \leq 13^\circ$ \\          \psfig{file=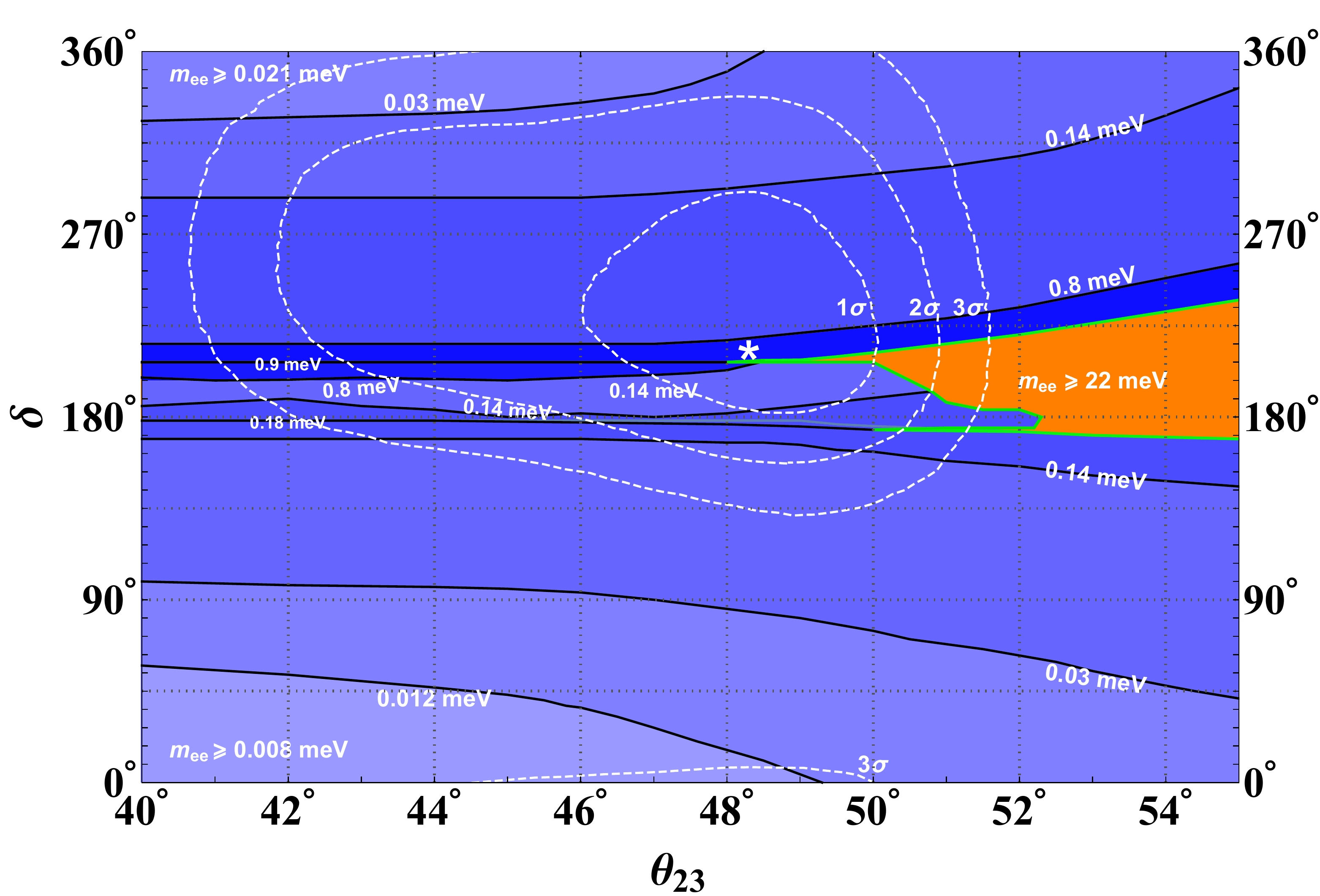,height=55mm,width=85mm}  \\ \vspace{5mm}
  $0\leq \theta_{ij}^L \leq 13^\circ$     \\  \psfig{file=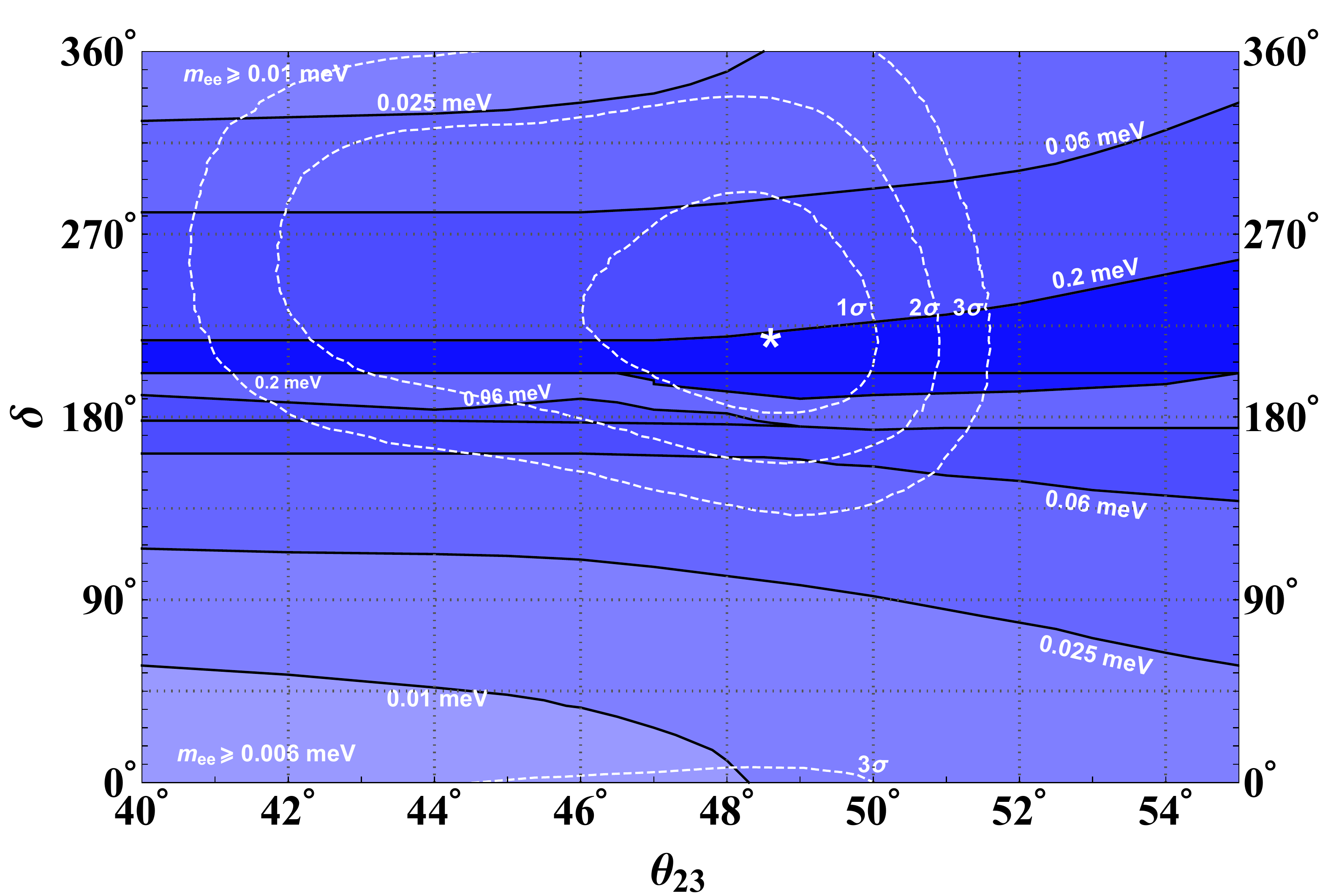,height=55mm,width=85mm}
\end{center}
\caption{Isocontour lines for the lower bound on $m_{ee}$ for the same three cases as in Fig.~5.}
\label{fig:VLmee}
\end{figure}
These results are somehow expected since allowing the angles in $V_L$ to vary freely, one has 
a full dependence of the asymmetry on six additional parameters and it is quite natural that the constraints on
low energy neutrino parameters would gradually disappear. {\em Our results indicate more clearly what size of the mixing angles
are needed to relax considerably the lower bound} (as we have seen at the level of one order of magnitude). This can be 
an indication for the identification of realistic fits within specific models that also aim at explaining the matter-antimatter
asymmetry with leptogenesis. 
We should also clarify that we selected only solutions that respect the condition $M_1 \lesssim 10^9\,{\rm GeV}$ for the applicability 
of the $N_2$ leptogenesis expression for the asymmetry Eq.~(\ref{twofl}). If one allows $M_1$ to become larger, then one can have
$N_1$ leptogenesis and the lower bound on the absolute neutrino mass scale would evaporate and in that case the
interference between just the two lightest RH neutrinos can be sufficient to reproduce the observed asymmetry, so that
the heaviest can be arbitrarily large and decouple from in the seesaw mechanism realising the two RH neutrino limit. 
From this point of view we can say that the value of the absolute neutrino mass scale sets a border between $N_1$-leptogenesis 
and $N_2$-leptogenesis.\footnote{However, even in the case when $N_1$ is heavier than $\sim 10^9\,{\rm GeV}$ one could have
regions in the space of parameters where the contribution from $N_2$ decays could be dominant and one would realise $N_2$ leptogenesis.
This possibility has been found even in the case of a two RH neutrino model \cite{twoRH,Samanta:2019yeg}.} 

\section{Analytical insight}    

We can now use the analytical description provided in \cite{decrypting} for $V_L = I$
and then extended in \cite{full} for $I \leq V_L \leq V_{CKM}$, to understand the results
obtained with the scatter plots and in particular the dependence of the $m_1$ lower bound 
on $\theta_{23}$ and $\delta$.

The analytical expression for $N_{B-L}^{\rm f}$ given in Section 2 can be simplified considering 
that only the tauon flavour asymmetry can reproduce the observed total asymmetry and, therefore, 
one can neglect the electron  and muon asymmetries and write from Eq.~(\ref{twofl})
\be\label{NBmLftau}
N_{B-L}^{\rm f} \simeq \ve_{2\tau} \, \kappa(K_{2\tau}) \, e^{-{3\pi \over 8}K_{1\tau}} \,  .
\ee

\subsection{Approximation $V_L = I$}

If we first consider, for simplicity, the limit $V_L = I$, then the three different 
quantities contributing to the final $B-L$ asymmetry, $\ve_{2\tau}$, $K_{1\t}$ and $K_{2\t}$, 
have the following simplified expressions just in terms of the nine low energy neutrino 
parameters  and $\a_2$ \cite{decrypting}
\be
\left.\ve_{2\tau}\right|_{V_L = I} = {3\over 16\,\pi}\, {\a_2^2\,m_c^2 \over v^2}\, {|m_{\nu ee}|\,
(|m^{-1}_{\nu \t \t}|^2 + |m^{-1}_{\nu \m \t}|^2)^{-1} \over m_1\,m_2\,m_3}
\, {|(m_{\nu}^{-1})_{\m \t}|^2 \over |(m_{\nu}^{-1})_{\t \t}|^2} 
\, \sin \a_L \,  ,
 \,   ,
\ee
\be
\left. K_{2\tau} \right|_{V_L = I} =
{m^2_{D3} \over m_{\star} \, M_2} \, |U_{R 3 2}|^2  
\simeq {m_1\,m_2\,m_3 \over m_{\star}}\, 
{|(m_{\nu}^{-1})_{\m \t}|^2 \over |m_{\nu ee}|\, |(m_{\nu}^{-1})_{\t \t}|} 
\ee
and
\be\label{K1tau}
\left. K_{1\tau} \right|_{V_L = I} =   
{m^2_{D3} \over m_{\star} \, M_1} \, |U_{R 3 1}|^2  
\simeq {|m_{\nu e\t}|^2 \over m_{\star}\,|m_{\nu ee}|}  
 \,    ,
\ee
where in the first expression we introduced 
the {\em effective $SO(10)$-inspired leptogenesis phase} 
\be
\a_L =  {\rm Arg}\left[m_{\nu ee}\right]  - 2\,{\rm Arg}[(m^{-1}_{\nu})_{\m\t}] - \pi -2\,(\rho+\s)  \,  .
\ee
Therefore, from Eq.~(\ref{NBmLftau})  we obtain an explicit expression for the final $B-L$ 
asymmetry in the approximation $V_L=I$ as a function just of $m_{\nu}$ and $\a_2$ \cite{decrypting}:
\bea\label{NBmLfVLI}
\left. N_{B-L}^{\rm lep,f} \right|_{V_L =I} & \simeq & 
{3\over 16\,\pi}\, {\a_2^2\,m_c^2 \over v^2}\, {|m_{\nu ee}|\,
(|m^{-1}_{\nu \t \t}|^2 + |m^{-1}_{\nu \m \t}|^2)^{-1} \over m_1\,m_2\,m_3}\,
{|m^{-1}_{\n\m\t}|^2\over |m^{-1}_{\n\t\t}|^2}\,\sin\a_L  \,  \times  \\  \nonumber
&  &\times  \; \kappa\left({m_1\,m_2\,m_3 \over m_{\star}}\, 
{|(m_{\nu}^{-1})_{\m \t}|^2 \over |m_{\nu ee}|\, |(m_{\nu}^{-1})_{\t \t}|} \right) \, \times \\  \nonumber
&  &\times \;
e^{-{3\pi\over 8}\,{|m_{\nu e\t}|^2 \over m_{\star}\,|m_{\nu ee}|}  }  
\,  .
\eea

From this analytical expression we can understand some of the numerical results we found,
though with some limitations due to the approximation $V_L = I$. In particular, we can understand the
lower bound on $m_1$ for  $\tau_A$ solutions.

\subsubsection{$\tau_A$ solutions}

The $\tau_A$ solutions are characterised by low values of $m_1$. For the derivation of the lower bound, 
we can safely specialise the expression (\ref{NBmLfVLI}) for the asymmetry in the hierarchical limit, 
for $m_1 \ll m_{\rm sol} \simeq 8.6\,{\rm meV}$. First of all it is important to  write the limit 
for the $N_1$ tauon flavoured decay parameter, since this describes the exponential wash-out.
In this limit, from Eq.~(\ref{K1tau}), one finds: 
\be
\left. K_{1\t} \right|_{V_L = I}(m_1 \ll m_{\rm sol}) \simeq 
{1\over m_{\star}}\,{|(m_{\rm atm}\,e^{i\,(2\s -\d)} -m_{\rm sol}\,s^2_{12}\,e^{i\,\delta})\,s_{13}\,c_{13}\,c_{23} - m_{\rm sol}\,c_{13}\,s_{12}\,c_{12}\,s_{23}|^2
\over |m_{\rm sol}\,s^2_{12}\,c_{13}^2 + m_{\rm atm}\,s^2_{13}\,e^{2i\,(\s -\d)}|}  \,  .
\ee
This expression can be recast conveniently in the following way 
\be\label{K1tauhier}
\left. K_{1\t} \right|_{V_L = I}(m_1 \ll m_{\rm sol}) \simeq 
s_{23}^2 \, c^2_{12} \, {m_{\rm sol}\over m_{\star}}\,{\left|{m_{\rm atm}\,e^{ i (2\s - \d)}-m_{\rm sol}\,e^{i \d}\,s_{12}^2 \over m_{\rm sol}}\,{s_{13}\over \tan\theta_{23}\,s_{12}\,c_{12}} - 1 \right|^2 \over 
\left|1 + {m_{\rm atm} \over m_{\rm sol}} {s^2_{13}\over c^2_{13}\,s^2_{12}}\,e^{2 i (\s - \d)}\right|}  \,  .
\ee
Currently, from reactor neutrino measurements, the mixing angle $\theta_{13}$ is known with 
great accuracy and precision (see Eq.~(\ref{theta13})). 
However, the expression (\ref{K1tauhier}) gives us the opportunity, as a side result,
also to highlight a successful feature of $SO(10)$-inspired leptogenesis. 
Considering that the dependence of $\ve_{2\tau}$ on $\theta_{13}$ can be neglected in first approximation 
and that $K_{2\tau}$ does not depend on $\theta_{13}$, 
one can derive both a lower bound and an upper bound on $\theta_{13}$ 
minimising $K_{1\tau}$ on the phases $\s$ and $\delta$ and imposing $K_{1\tau} \lesssim 1$. 
First of all the possibility of a cancellation in the numerator of (\ref{K1tauhier}) is possible
if $2\s-\delta = 2 \,n \, \pi$ with $n$ integer in a way that $\exp[i (2\s - \d)]=1$.
This is a condition that needs to be realised quite strictly and indeed in the scatter plots 
the quantity $2\s -\d$ is observed to peak narrowly around values $2\,n \, \pi$.
Secondarily, one has also to impose $\exp[2 i (\s - \d)]=1$ implying $\s - \d = m\,\pi$.
As we will see, this latter condition is also necessary to maximise the $C\!P$ asymmetry.
Imposing both conditions simultaneously, one has $\s \simeq m\,\pi$ 
and $\delta = 2\, n \, \pi$. These different periodicities for $\s$ and $\d$ are clearly observed in scatter plots. 
We can then write 
\be\label{K1taumin}
K_{1\tau} \gtrsim K_{1 \tau}^{\rm min} \equiv  
s_{23}^2 \, c^2_{12} \, {m_{\rm sol}\over m_{\star}}\,{\left({m_{\rm atm}-m_{\rm sol}\,s_{12}^2 \over m_{\rm sol}}\,{s_{13}\over \tan\theta_{23}\,s_{12}\,c_{12}} - 1 \right)^2 \over 
1 + {m_{\rm atm} \over m_{\rm sol}} {s^2_{13}\over c^2_{13}\,s^2_{12}}}  \,  .
\ee
Let us now focus on the dependence of $K_{1 \tau}^{\rm min}$ on $\theta_{13}$. 
First, notice that $s_{23}^2 \, c^2_{12} \, {m_{\rm sol} / m_{\star}} \simeq 3$ and, in the limit $s_{13} \ra 0$, 
one would have an exponential suppression not compatible with successful leptogenesis.  In this limit
the term proportional to $s_{13}$ in the numerator of Eq.~(\ref{K1tauhier}) is smaller than $1$, while the 
term $\propto s^2_{13}$ in the denominator  can be neglected.
Imposing $K_{1\tau}^{\rm min} \lesssim 1$, then implies the lower bound \cite{decrypting}
\be
s_{13} \gtrsim \tan \theta_{23}\,s_{12}\,c_{12}\,{m_{\rm sol}\over m_{\rm atm}} \,
\left(1 - {1\over s_{23}\, c_{12} \sqrt{m_{\rm sol}/m_{\star}}}\right) \simeq 0.03\, \tan\theta_{23} \,  ,
\ee
corresponding to $\theta_{13} \gtrsim 2^{\circ}$ for $\theta_{23} \gtrsim 41^{\circ}$. 
On the other hand, when the term $\propto s_{13}$ is larger than unity, 
retaining the small correcting term $\propto s^2_{13}$ in the denominator, one obtains an upper bound
\be
s_{13} \lesssim \tan \theta_{23}\,s_{12}\,c_{12}\,{m_{\rm sol}\over m_{\rm atm}} \, 
\left[1 + {1\over s_{23}\, c_{12} \sqrt{m_{\rm sol}/m_{\star}} }\,
\left(1 + {1\over 2}\,{m_{\rm atm} \over m_{\rm sol}}\, {s^2_{13} \over c^2_{13}\,s^2_{12} } \right)
\right] \simeq 0.145 \, \tan\theta_{23} \,  ,
\ee
giving $\theta_{13} \lesssim 10.3^{\circ}$ for $\theta_{23} \lesssim 51^{\circ}$ (see Eq.~(\ref{theta23})). 
The found allowed range, $2^\circ \lesssim \theta_{13} \lesssim 10.3^\circ$ for $V_L = I$, nicely reproduces
the numerical results (see for example \cite{decrypting}).

Let us now derive the lower bound on $m_1$. In this case we also need to consider the limit of $\ve_{2\tau}$
for $m_1 \ll m_{\rm sol}$, obtaining \cite{decrypting}
\be
\left.\ve_{2\t}\right|_{V_L = I}(m_1 \ll m_{\rm sol})  \simeq
{3\over 16\,\pi}\,{\a_2^2 \, m_c^2 \over v^2} \, 
{m_1 \over m_{\rm sol}\,m_{\rm atm}} \, {|m_{\rm sol}\,U_{e2}^2+m_{\rm atm}\,U^2_{e3}| \, |U_{\m 1}|^2\over 
|U_{\t 1}|^4\, (|U_{\t 1}|^2 + |U_{\m 1}|^2)} \,
\sin \a_L \,  ,
\ee
with $\a_L(m_1 \ll m_{\rm sol}) \simeq 2\,(\r - \s) $. 
Clearly this is maximised for $\sin\a_L = 1$ implying $\r = \pi/4 + \s + n \pi$.\footnote{Though notice that this is not the 
condition maximising $N_{B-L}^{\rm f}$
 as explained in \cite{decrypting}, since $\rho$ also appears in $K_{1\tau}$ if one takes
 into account a sub-dominant term $\propto m_1$ and this shifts the condition for maximising the asymmetry
 from $\rho = \pi/4 + \s + n\,\pi$ to $\rho \simeq 0.35\,\pi + \s + n\pi$. In any case there is a phase difference
 between $\rho$ and $\sigma$.} Retaining terms proportional to $s_{13}$ in $|U_{\tau 1}|$ and $|U_{\mu 1}|$ 
 (they were neglected in \cite{decrypting}),  we can then write
\bea\label{ve2tauVL1}
\left.\ve_{2\t}\right|_{V_L = I}(m_1 \ll m_{\rm sol})  & \lesssim &
{3\over 16\,\pi}\,{\a_2^2 \, m_c^2 \over v^2} \, 
{m_1 \over m_{\rm sol}\,m_{\rm atm}} \, \times \\  \nonumber
&  & \times \, {|m_{\rm sol}\,s^2_{12}\,c^2_{13}
+m_{\rm atm}\,s^2_{13}\,e^{2\,i\,(\s-\d)}|\,|s_{12}\,c_{23}+c_{12}\,s_{23}\,s_{13}\,e^{i\d}|^2 \over |s_{12}\,s_{23}-c_{12}\,c_{23}\,s_{13}e^{i\d}|^4 \, s^2_{12} } \,   ,
\eea
where we used $|U_{\t 1}|^2 + |U_{\m 1}|^2 \simeq s^2_{12}$.
 Notice that this expression is maximised for $\s - \d \simeq n\, \pi$ and $\d = 2\,m\,\pi$ with $n,m$ integers, 
 the same conditions that were minimising $K_{1\tau}$.\footnote{Notice 
that $\ve_{2\tau} \propto m_1$. This comes from the fact that the $C\!P$ asymmetry 
is generated by the  interference of $N_2$-decays with $N_3$ in the loops. 
Since $M_3 \propto m_1^{-1}$, the limit $m_1 \ra 0$ corresponds to the limit when $N_3$
decouples and the  interference, encoded by ${\cal I}^\tau_{23}$ vanishes: 
this is the physical origin of the lower bound on $m_1$.}
We can then write
 \be
 \left.\ve_{2\t}\right|_{V_L = I}(m_1 \ll m_{\rm sol})   \lesssim 
 {3\over 16\,\pi}\,{\a_2^2 \, m_c^2 \over v^2} \, 
\, {m_1 \over m_{\rm atm}} 
{c^2_{13}\,(s_{12}\,c_{23}+c_{12}\,s_{23}\,s_{13})^2\over (s_{12}\,s_{23}-c_{12}\,c_{23}\,s_{13})^4}\, \left(1+{m_{\rm atm}\,s^2_{13}\over m_{\rm sol}\,s^2_{12}\,c^2_{13}} \right) \,  .
 \ee
 The third and last ingredient to consider to maximise the final $B-L$ asymmetry 
 and calculate the lower bound on $m_1$ is the efficiency factor at the production $\kappa(K_{2\tau})$. 
In the limit $m_1 \ll m_{\rm sol}$, the flavoured decay parameter
 \be
\left. K_{2\tau} \right|_{V_L =I}(m_1 \ll m_{\rm sol}) = 
c^2_{23}\,{m_{\rm atm} \over m_{\star}} \gtrsim 20 \,  \;  .
 \ee
This shows that $\tau_A$ solutions are characterised by strong wash-out at the production 
and in this case one can use approximately \cite{flavourlep}
\be
\kappa(K_{2\tau} \gg 1) \simeq {0.5 \over K_{2\tau}^{1.2}} \simeq {0.5 \over c_{23}^{2.4}} \, 
\left({m_\star \over m_{\rm atm}} \right)^{1.2} \,  .
\ee
Finally, following Eq.~(\ref{NBmLftau}), we can put all three terms together and write
\be\label{NBmLfub}
N_{B-L}^{\rm f} \lesssim  {3\over 32\,\pi}\,{\a_2^2 \, m_c^2 \over v^2} \, 
\, {m_1 \over m_{\rm atm}} \, \left({m_\star \over m_{\rm atm}} \right)^{1.2} 
{c^2_{13}\,(s_{12}\,c_{23}+c_{12}\,s_{23}\,s_{13})^2\over c_{23}^{2.4} \, (s_{12}\,s_{23}-c_{12}\,c_{23}\,s_{13})^4}\, \left(1+{m_{\rm atm}\,s^2_{13}\over m_{\rm sol}\,s^2_{12}\,c^2_{13}} \right) \,e^{-{3\pi \over 8}K_{1\tau}^{\rm min}} \,  ,
\ee 
where $K_{1\tau}^{\rm min}$ is given by the expression (\ref{K1taumin}).  From $N_{B-L}^{\rm f}$ one can 
then obtain $\eta_B^{\rm lep}$ using simply Eq.~(\ref{etaBlep}) and, imposing $\eta_B^{\rm lep} = \eta_B^{\rm exp}$ 
(see Eq.~(\ref{etaBexp})), from the upper bound (\ref{NBmLfub}) one finally obtains the lower bound
\bea\label{m1lb}
m_1 & \gtrsim & m_{\rm atm} \, {32\,\pi \over 3}\, {\eta_B^{\rm exp} \over 0.96 \times 10^{-2}} \, {v^2 \over \a_2^2 \, m_c^2} \, 
\, \left({m_{\rm atm}  \over m_\star} \right)^{1.2} 
{c_{23}^{2.4} \, s_{12}^4\,s_{23}^4 \, (1- {c_{12}\,c_{23}\,s_{13} \over s_{12} \, s_{23}})^4 \over c^2_{13}\,(s_{12}\,c_{23}+c_{12}\,s_{23}\,s_{13})^2}\, \times  \\ \nonumber
&  & \hspace{70mm}\times \left(1+{m_{\rm atm}\,s^2_{13}\over m_{\rm sol}\,s^2_{12}\,c^2_{13}} \right)^{-1} \,e^{{3\pi \over 8}K_{1\tau}^{\rm min}} \,  ,
\eea 
that we have recast in a way to highlight that $m_1^{\rm min}\propto s_{23}^4$.
If we now use the experimental values for $\theta_{13}$, $\theta_{12}$, $m_{\rm atm}$, $m_{\rm sol}$ and $\eta_B^{\rm exp}$,
we obtain a lower bound on $m_1$ depending just on $\theta_{23}$.
Despite the fourth power dependence on $s_{23}$, that would tend to make the lower bound more stringent at higher values
of $\theta_{23}$, the dependence on $\theta_{23}$ in $K_{1\tau}^{\rm min}$ is stronger and actually the lower bound gets more
relaxed for increasing values of $\theta_{23}$. For example, one finds for the 
$3\s$ extreme allowed $\theta_{23}$ values and for the best fit value: 
\bea\label{lbnum}
\theta_{23}  & = &  40. 8^{\circ} \Rightarrow K_{1\tau}^{\rm min}\simeq  1.87 \, , \; m_1 \gtrsim 5\,{\rm meV} \,  , \\  \nonumber 
\theta_{23}  & = & 48.6^\circ  \Rightarrow K_{1\tau}^{\rm min}\simeq 0.71 \, , \;  m_1    \gtrsim  3\,{\rm meV} \,  , \\ \nonumber
\theta_{23}  & = & 51.3^\circ \Rightarrow K_{1\tau}^{\rm min}\simeq  0.43 \, , \;  m_1 \gtrsim 2.5\,{\rm meV}   \,   .
\eea
These results are in  good agreement with the numerical results found in \cite{decrypting,full},
they just overestimate the lower bound by $ \sim 1\,{\rm meV}$, a discrepancy that 
would be fully corrected if one would include the sub-dominant term $\propto m_1$ 
in the expression for $K_{1\tau}$ and then finding the lower bound on $m_1$ solving by iteration.
However, the explicit expression we obtained well describes the dependence on $\theta_{23}$ in the case $V_L = I$.

We can also understand the dependence on $\delta$. Since the condition $2\s - \d = 2 n \pi $ needs to be verified
in a stringent way we can then rewrite $2(\s - \d) = -\delta +2 n \pi$ both in the expression for
$K_{1\tau}$ (see Eq.~(\ref{K1tauhier})) and in that one for $\ve_{2\tau}$ (see Eq.~(\ref{ve2tauVL1})),
obtaining
\bea\label{m1lbd}
m_1 & \gtrsim & m_{\rm atm} \, {32\,\pi \over 3}\, {\eta_B^{\rm exp} \over 0.96 \times 10^{-2}} \, {v^2 \over \a_2^2 \, m_c^2} \, 
\, \left({m_{\rm atm}  \over m_\star} \right)^{1.2} 
{c_{23}^{2.4} \, s_{12}^4\,s_{23}^4 \, \left|1- {c_{12}\,c_{23}\,s_{13} \over s_{12} \, s_{23}}\, e^{i\d}\right|^4 \over c^2_{13}\,
|s_{12}\,c_{23}+c_{12}\,s_{23}\,s_{13}\, e^{i\d}|^2}\, \times  \\ \nonumber
&  & \hspace{70mm}\times \left|1+{m_{\rm atm}\,s^2_{13}\over m_{\rm sol}\,s^2_{12}\,c^2_{13}}\, e^{-i\d} \right|^{-1} \,e^{{3\pi \over 8}K_{1\tau}^{\rm min}(\d)} \,  ,
\eea 
where
\be\label{K1taumind}
K_{1 \tau}^{\rm min}(\d) \equiv  
s_{23}^2 \, c^2_{12} \, {m_{\rm sol}\over m_{\star}}\,{\left({m_{\rm atm}-m_{\rm sol}\,s_{12}^2 \over m_{\rm sol}+m_1}\,{s_{13}\over \tan\theta_{23}\,s_{12}\,c_{12}} - 1 \right)^2 \over 
\left|1 + {m_{\rm atm} \over m_{\rm sol}} {s^2_{13}\over c^2_{13}\,s^2_{12}}\, e^{-i\d} \right|}  \,  .
\ee
Notice that in this case the lower bound becomes very close to $m_{\rm sol}$ and we included
a term proportional to $m_1$. This implies that the lower bound now is not in an explicit form and has to be
solved iteratively. Eq.~(\ref{K1taumind})  reproduces well the effect of $\delta$ in increasing
$K_{1\t}^{\rm min}(\d)$ making the lower bound more stringent. This expression gives good results for $|\delta|< \pi/2 $, 
for higher values the lower bound becomes close to $m_{\rm sol}$ and one has to use the full expression. 
Moreover, there are two critical values of $\delta$, one below $\pi$ and one higher, for which the lower bound 
becomes equal to the upper bound and the allowed $m_1$ range closes up. Therefore, 
between these two critical values there are no $\tau_A$ solutions and this well explains what observed in the scatter plots.
Within this $\delta$ window, the lower bound on $m_1$ has to be calculated within the region of $\tau_B$ solutions.

\subsubsection{$\tau_B$ solutions}

A more detailed discussion on $\tau_B$ solutions can be found in \cite{decrypting}. Here we just recall that for these solutions,
since $m_1 \gg m_{\rm sol} \sim 10\,{\rm meV}$, one can use the approximation $m_1 \simeq m_2$. In this case from
the full expression of $K_{1\tau}$ one finds that in order for this to be not too large, one needs $\rho \simeq n\,\pi$. 
This immediately allows to understand why for $\tau_B$ solutions one necessarily has $m_{ee} \simeq m_1$.
From Eq.~(\ref{mee}) one can write explicitly  
\be
m_{ee} = \left|m_1\, c^2_{12}\,c^2_{13}\,e^{2i\rho}+m_2\, s^2_{12} \, c^2_{13} +m_3\,s^2_{13}\,e^{2i\,(\s -\d)} \right| \,   .
\ee
Therefore, for $\tau_B$ solutions one immediately finds 
\be
m_{ee} \simeq \left| m_1 \, c^2_{13} + m_3\,s^2_{13}\,e^{2i (\s -\d}) \right| \,  ,
\ee
showing that for $\tau_B$ solutions necessarily $m_{ee} \simeq m_1$ and more precisely
one has $m_{ee} < m_1$ with $|m_{ee}-m_1|< 2\,m_3\,s^2_{13}$, quite a distinctive feature that can 
be regarded as a signature of $\tau_B$ solutions.  
Like for $\tau_A$ solutions, also for $\tau_B$ solutions there are  both a lower bound and an upper bound on $m_1$.
They both depend on $\theta_{23}$ in a way that the interval gradually shrinks
for increasing values of $\theta_{23}$ up to a maximum value of $\theta_{23}$, well above
the experimentally allowed range, where the interval closes up \cite{decrypting}.

The range of allowed $m_1$ values, using the approximation $m_1 \simeq m_2 \gg m_{\rm sol}$, is 
approximately determined by imposing $\left.\eta_B^{\rm lep}\right|_{V_L = I}(m_1 \gg m_{\rm sol}) = \eta_B^{\rm exp}$ where 
\bea 
\left.\eta_B^{\rm lep,max}\right|_{V_L = I}(m_1 \simeq m_2 \gg m_{\rm sol})  &\simeq & 
0.96 \times 10^{-2} \, {3 \over 16 \, \pi} \, {\a_2^2 \, m_c^2 \over v^2} \, \times  \\ \nonumber
& & \times {m_1 \over m_3} \, {s^2_{23}\, c^2_{23}\, (1-{m_1 \over m_3}\,c^2_{13})^2/(s^2_{23}+{m_1 \over m_3}\,c^2_{23}\,c^2_{13})^2
\over (s^2_{23}\, c^2_{23}\, (1-{m_1 \over m_3}\,c^2_{13})^2+s^2_{23}+{m_1 \over m_3}\,c^2_{23}\,c^2_{13})} \, \times \\ \nonumber
 &  &   \times \kappa(K_{2\tau}^{\rm min}) \, e^{{3\pi \over 8}K_{1\tau}^{\rm min}} \,   ,   
\eea
with 
\be
K_{1\tau}^{\rm min} = s^2_{13}\,c^2_{13} \, c^2_{23}\, {(m_3 -m_1)^2 \over m_\star \, (m_1 + m_3 \,s^2_{13})} \,    \;\; \mbox{\rm and} \;\;
K_{2\tau}^{\rm min} = {m_3 \over m_\star}\,{s^2_{23}\,c^2_{23}\,(1-{m_1 \over m_3}\,c^2_{13})^2 \over \left(s^2_{23}+{m_1 \over m_3}\,c^2_{23}\right)^2} \,  .
\ee
Notice that the asymmetry is maximised for $\rho \simeq n\, \pi$, $\d \simeq m \pi$ and $\s \simeq -\pi/8 +k\,\pi$ (with $n,m,k$ integers). Imposing 
$\left.\eta_B^{\rm lep,max}\right|_{V_L = I}(m_1 \simeq m_2 \gg m_{\rm sol})  \geq n_B^{\rm exp}$ one can find a range of allowed values for $m_1$
as a function of $\theta_{23}$. For example, for the 
$3\s$ extreme allowed $\theta_{23}$ values and for the best fit value the following ranges, one finds:
\bea\label{lbnum}
\theta_{23}  & = &  40. 8^{\circ} \Rightarrow   35 \, {\rm meV} \lesssim m_1 \lesssim 70\,{\rm meV} \,  , \\  \nonumber 
\theta_{23}  & = & 48.6^\circ  \Rightarrow       41 \, {\rm meV} \lesssim m_1 \lesssim 65\,{\rm meV} \,  , \\ \nonumber
\theta_{23}  & = & 51.3^\circ \Rightarrow         42 \, {\rm meV} \lesssim m_1 \lesssim 62\,{\rm meV}    \,   .
\eea

 \subsection{$I \leq V_L \leq V_{CKM}$}    

The account of a small mismatch between neutrino Dirac mass matrix and charged lepton mass matrix
comparable to the same mismatch observed in the quark sector between up-quark and down-quark mass matrices,
is encoded by the expressions (\ref{ve2alANtau})--(\ref{KialVL}) and was studied in detail in \cite{full}. Here we want to specialise some 
of the analytical considerations made in \cite{full} to the lower bound on the absolute neutrino mass scale.

The tauon flavoured asymmetry $\ve_{2\tau}$ gets slightly corrected by turning on small mixing angles in $V_L$. 
The same it is true for $K_{2\tau}$ and consequently the wash-out factor $\kappa(K_{2\tau})$. The quantity that is very sensitive to
a small deviation of $V_L$ from the identity is the $N_1$ wash-out factor since this is an exponential and the argument 
is proportional to $K_{1\tau}$. As one can see from Eq.~(\ref{K1taumin}), for $V_L = I$ this is proportional to the square 
of the difference of two quantities both close to unity. This difference is sensitive to $\theta_{23}$ 
and, in particular, for values close to the lower  edge of the experimental $3\s$ range, there is no cancellation and one has $K_{1\tau} \simeq 2$ (see Eq.~\ref{lbnum})) so that the wash-out suppression is quite large. However, when $V_L \simeq V_{CKM}$, then new terms enter the expression for $K_{1\tau}$ and one can have $K^{\rm min}_{1\tau} \ll 1$ in the Eq.~(\ref{m1lb}) independently of $\theta_{23}$.  
Let us see this result explicitly. From Eqs.~(\ref{MI}), (\ref{UR}) and (\ref{KialVL}), one can derive the expression \cite{full}
\be\label{K1tau}
K_{1\t} \simeq 
{1\over m_{\star}}\,
\left( {|\widetilde{m}_{\nu 13}|^2 \over |\widetilde{m}_{\nu 11}|}\,|V_{L33}|^2
+2\,{V_{L23}\,V_{L33}^{\star} \over |\widetilde{m}_{\nu 11}|}\,
{\rm Re}\left[\widetilde{m}_{\nu 12}^\star \, \widetilde{m}_{\nu 13} \right]
+|V_{L 23}|^2 \, {|\widetilde{m}_{\nu 13}|^2 \over |\widetilde{m}_{\nu 11}|} \right) \,  .
\ee
The first term is the dominant one and if we
choose $\theta^L_{13}= \theta^L_{23} = 0$ the others vanish exactly.
From the definition $\widetilde{m}_{\nu} = V_L\,m_{\nu}\,V_L^T$ and from the 
parameterisation of $V_L$, Eq.~(\ref{VLmatrix}), we arrive to the following
expression for $K_{1\tau}^{\rm min}$
\be\label{K1tauminVL}
K_{1\tau} \gtrsim K_{1 \tau}^{\rm min}(\theta_{12}^L) \equiv  
{m_{\rm sol}\over m_{\star}}\,{\left[s_{23}^2 \, c^2_{12} \, \left({m_{\rm atm}-m_{\rm sol}\,s_{12}^2 \over m_{\rm sol}}\,{s_{13}\over \tan\theta_{23}\,s_{12}\,c_{12}} - 1 \right)e^{i\s_L}+{1\over 2}\,\sin 2\theta_{23}\,\sin\theta_{12}^L \,{m_{\rm atm} \over m_{\rm sol}} \right]^2 \over 
1 + {m_{\rm atm} \over m_{\rm sol}} {s^2_{13}\over c^2_{13}\,s^2_{12}}}  \,  .
\ee
This expression clearly shows that for $\s_L \simeq (2n+1)\,\pi$ one can have a cancellation for any value of $\theta_{23}$
for a proper value of $\theta_{12}^L$. In this way the wash-out from $N_1$ inverse processes can always be suppressed.

This is the dominant effect of accounting for $V_L \simeq V_{CKM}$ and that makes in a way that 
one can always find solutions with negligible lightest RH neutrino wash-out. 
In this way the  dependence of the $m_1$ lower bound  on $\theta_{23}$ coming from $K_{1\tau}^{\rm min}$ disappears
and one is left only with the dependence  from $\ve_{2\tau}\propto s_{23}^{-4}$ so that the
lower bound now gets relaxed for decreasing values of $\theta_{23}$ and the 
numerical values for the lower bound reported  in Eq.~(\ref{lbnum}) relax into
\bea\label{lbnumVL}
\theta_{23}  & = &  40. 8^{\circ} \Rightarrow   m_1 \gtrsim 0.6\,{\rm meV} \,  , \\  \nonumber 
\theta_{23}  & = &  48.6^\circ    \Rightarrow    m_1    \gtrsim  1.3\,{\rm meV} \,  , \\ \nonumber
\theta_{23}  & = &  51.3^\circ    \Rightarrow     m_1 \gtrsim 1.5\,{\rm meV}   \,   ,
\eea
very well reproducing the numerical results. There is also some 
relaxation of about $10\,{\rm meV}$ of the lower bound on $m_1$ for $\tau_B$ solutions, 
that however we do not discuss.

\section{Final remarks}    

The $SO(10)$-inspired leptogenesis scenario is a remarkable example of how also high energy
scale leptogenesis models are testable when a proper reduction of the number of independent 
parameters is realised imposing additional conditions. The latest results from neutrino oscillation experiments have
started favouring a region in the plane $\delta$ and $\theta_{23}$
that allows to establish a very interesting connection between the absolute neutrino 
mass scale and mixing parameters. We have seen that essentially for the large values of $\theta_{23}$,
now favoured by neutrino oscillation experiments, there are two well distinguished allowed regions: one at low values 
and one at high values of $m_1$. If current best fit values will be confirmed by next generation of long baseline
experiments, DUNE and T2HK, confirming moreover discovery of $C\!P$ violation, 
then $SO(10)$-inspired leptogenesis would favour $m_1 \gtrsim 34\,{\rm meV}$ and $m_{ee}\gtrsim 31\,{\rm meV}$,
implying that absolute neutrino mass scale experiments should either find a signal during next years, both 
from cosmology and from neutrinoless double beta decay experiments with $m_{ee} \simeq m_1$, or place a severe constraint
$\a_2 \gtrsim 5$. It is of course particularly exciting that, despite neutrino masses are normally ordered, 
neutrinoless double beta decays should be observed, implying a discovery of lepton number violation.   
On the other hand, if in the end the value of $\delta$
should lie in the fourth quadrant, rather than in the third one, and  $\theta_{23}$ will be confirmed in the second octant, 
the range of allowed values of $m_1$ within $SO(10)$-inspired leptogenesis would be approximately $(2$--$10)\,{\rm meV}$, 
still partly testable by cosmological observations if current tensions 
in the $\L$CDM model are solved and  a sensitivity to a departure from the hierarchical limit in the
sum of the neutrino masses at the level of the value of $m_1$ can be reached.  
We are then entering an exciting experimental phase when low energy neutrino experiments are  
effectively testing an attractive scenario for the origin of the matter-antimatter asymmetry of the universe 
that emerges within a class of models typically realised within grand-unified theories: it is certainly
a timely opportunity for $SO(10)$-inspired leptogenesis.

\vspace{-1mm}
\subsection*{Acknowledgments}

PDB  acknowledges financial support from the STFC Consolidated Grant L000296/1. 
RS is supported  by a Newton International Fellowship (NF 171202) from Royal Society (UK) and SERB (India). 
This project has received funding/support from the European Union Horizon 2020 research and innovation 
programme under the Marie Sk\l{}odowska-Curie  grant agreements number 690575 and  674896.

\end{document}